\documentclass[letterpaper]{JHEP3}
\usepackage{amsmath}
\usepackage{dsfont}
\usepackage{amssymb}
\usepackage{bm}
\newcommand{\eq}{\begin{equation}}
\newcommand{\eeq}{\end{equation}}

\title{A Cardy-like formula for rotating black holes with planar horizon}

\author{Mois\'es Bravo Gaete\\
   Facultad de Ciencias B\'asicas, Universidad
Cat\'olica del Maule, Casilla 617, Talca, Chile.\\
    E-mail: \email{mbravo-at-ucm.cl}}

\author{Luis Guajardo\\
Instituto de Matem\'atica y Fisica, Universidad de Talca,
Casilla 747, Talca, Chile.\\
E-mail: \email{luis.guajardo.r-at-gmail.com}}

\author{Mokhtar Hassa\"ine\\
Instituto de Matem\'atica y Fisica, Universidad de Talca,
Casilla 747, Talca, Chile.\\
E-mail: \email{hassaine-at-inst-mat.utalca.cl}}

\abstract{We show that the semiclassical entropy of $D-$dimensional
rotating (an)isotropic black holes  with planar horizon can be
successfully computed according to a Cardy-like formula. This
formula does not refer to any central charges but instead involves
the vacuum energy which is identified with a gravitational bulk
soliton. The soliton is obtained from the non-rotating black hole
solution by means of a double analytic continuation. The robustness
of the Cardy-like formula is tested with numerous and varied
examples, including AdS, Lifshitz and hyperscaling violation planar
black holes.}

\begin{document}
\section{Introduction}
Since the seminal works of Bekenstein and Hawking
\cite{Bekenstein:1973ur,Hawking:1974sw}, black holes are believed to
behave as thermodynamic objects with characteristic temperature and
entropy. A natural question has then emerged concerning the
statistical interpretation of the black hole entropy. One of the
first results that has shed some light on this problem was the
observation that the asymptotic symmetries of the three-dimensional
AdS space consist in two copies of the Virasoro algebra with a
central charge \cite{Brown:1986nw}. This latter corresponds to the
symmetry group of a two-dimensional CFT and, in this case, the Cardy
formula is well-appropriate to express the asymptotic density of
states \cite{Cardy:1986ie}. An important manifestation of the
AdS/CFT correspondence was then provided by showing that the Cardy
formula applied for the BTZ black hole \cite{Banados:1992wn}
correctly reproduces the expression of the Bekenstein-Hawking
entropy \cite{Strominger:1997eq}. Soon after, this approach was
generalized for higher-dimensional black holes having a
two-dimensional CFT dual in the case of standard General Relativity
\cite{Guica:2008mu} as well as in presence of higher-derivative
corrections \cite{Azeyanagi:2009wf}.

Extensions of the Cardy formula have been considered and studied in
the current literature. Among other, one can mention the
higher-dimensional generalization of the Cardy formula which applied
for strongly coupled field theories having an AdS dual
\cite{Verlinde:2000wg}. There also exist extensions of the Cardy
formula with applications for three-dimensional spacetimes that are
not AdS like the warped AdS spaces \cite{Detournay:2016gao} or
anisotropic spacetimes, namely the Lifshitz spacetimes
\cite{Gonzalez:2011nz} or the hyperscaling violation geometries
\cite{Shaghoulian:2015dwa,Bravo-Gaete:2015wua}. The interest on
anisotropic spacetimes has considerably grown up this  last decade
essentially due to the will of extending the ideas underlying the
gauge/gravity duality to strongly coupled field theories with an
anisotropic scaling symmetry \cite{Kachru:2008yh}. Notice also that
the cases of three-dimensional black holes have  permitted a better
comprehension of the holographic derivation of the entropy by
highlighting the prominent role played by the soliton, see Refs.
\cite{Correa:2010hf,Correa:2011dt,Correa:2012rc}. This observation
is not in contradiction with the standard derivation of the Cardy
formula for which the ground state is implicitly assumed to be the
three-dimensional AdS spacetime. Nevertheless, this assumption is in
general valid only for the vacuum sector but not in the hairy sector
which possesses a different ground state. It is then more judicious
to deal with a Cardy formula written in terms of the vacuum energy
rather than the central charges. Unfortunately, it is not possible
{\it a priori} to find out the vacuum energy of the putative field
theory. However, as stressed in Refs.
\cite{Correa:2010hf,Correa:2011dt,Correa:2012rc}, the vacuum energy
can be identified with the mass of a bulk soliton constructed from
the black hole through a double Wick rotation in the same way that
the AdS soliton \cite{Horowitz:1998ha}, which reinforces the
importance of the role played by the soliton. Importance also
confirmed  in the Lifshitz case \cite{Gonzalez:2011nz} where the
robustness of the Lifshitz Cardy formula has been tested
successfully for three-dimensional Lifshitz black holes with a
nonminimally scalar field \cite{Ayon-Beato:2015jga}.

One of the aim of this paper is precisely to confirm the importance
of the gravitational soliton. We will highlight this importance in
the case of rotating black holes in arbitrary dimension $D$ with a
planar base manifold. This restriction on the horizon's topology is
justified by the fact that the soliton can be easily constructed
from the black hole by a double analytic continuation similar to the
one operated in the AdS soliton \cite{Horowitz:1998ha}. With the
view of achieving this task, we will be interested on
higher-dimensional extensions of the Cardy formula for field
theories satisfying the following two assumptions: (i) the field
theory possibly displays a hyperscaling violation reflected by the
fact that the thermal entropy ${\cal S}$ scales with respect to the
temperature as ${\cal S}\sim T^{\frac{d_{\tiny{\mbox{eff}}}}{z}}$,
and (ii) the ground state for the field theory is identified with a
bulk soliton which is regular everywhere and devoid of any
integration constant. Here, $d_{\tiny{\mbox{eff}}}$ is an effective
spatial dimensionality for the dual theory (related to the dimension
of the stress-energy tensor) which measures the possible deviation
from the spatial dimension and $z$ is the Lifshitz dynamical
exponent. In the standard AdS situation, the "effective" spatial
dimension $d_{\tiny{\mbox{eff}}}=D-2$ and the dynamical exponent
$z=1$. Under these hypothesis, formulas for the asymptotic growth of
the number of states have been obtained in the non-rotating case in
\cite{Shaghoulian:2015dwa,Bravo-Gaete:2015wua}, and their spinning
generalizations in the isotropic case $z=1$ were found in
\cite{Shaghoulian:2015lcn}. In the present work, we extend this
formula for a generic dynamical exponent $z$. The resulting
Cardy-like formula makes no mention to any central charge but
instead involves the mass of the ground state which is identified
with the nonrotating gravitational soliton. Nevertheless, since a
very little is known about the putative field theories, we propose
to corroborate the validity of the Cardy-like formula considering
gravity theories whose spectrum of solutions contain black holes
whose entropy exhibits a power law temperature as described in the
hypothesis (i) as well as regular solitons (ii). This inspection
will be done for different classes of black hole solutions with
different asymptotic behaviors (rotating AdS, Lifshitz and
hyperscaling violation black holes with a planar base manifold) by
comparing the gravitational entropy  with the entropy field
expression involving the effective spatial dimension and the vacuum
energy. In all our examples, the spinning planar black holes are
derived from static configurations through a Lorentz boost and the
vacuum energy corresponds to the mass of the gravitational soliton
obtained from the nonrotating black hole through a double Wick
rotation. We will also extend these results to the case of charged
planar black holes, where again the ground state is identified with
the soliton derived from the neutral and nonrotating black hole.
Since the soliton is devoid of any integration constant, its mass
will be computed using the  quasilocal generalization of the ADT
formalism \cite{Abbott:1981ff} as presented in Refs.
\cite{Kim:2013zha,Gim:2014nba}. One of the main result of these two
last papers lies in the prescription of the off-shell ADT potential
${\cal Q}_{\mathrm{ADT}}^{\mu\nu}$ in terms of the off-shell Noether
potential $K^{\mu\nu}$ and the surface term $\Theta^{\mu}$ arising
from the variation of the action
\begin{equation}
\sqrt{-g}{\cal Q}_{\mathrm{ADT}}^{\mu\nu}= \frac{1}{2}\delta
K^{\mu\nu}-\xi^{[\mu}\Theta^{\nu]},
\end{equation}
where $\xi^{\mu}$ denotes the Killing vector. The corresponding
conserved charge is computed to be
\begin{equation}\label{quasilocalcharge}
Q(\xi)=\int\!d^{D-2}x_{\mu\nu}
\Big(\Delta{K}^{\mu\nu}(\xi)-2\xi^{[\mu}\!\!
\int^1_0\!\!ds~\Theta^{\nu]}(\xi|s)\Big),
\end{equation}
where $\Delta{K}^{\mu\nu}(\xi)\equiv
{K}^{\mu\nu}_{s=1}(\xi)-K^{\mu\nu}_{s=0}(\xi)$ denotes the
difference of the Noether potential between the black hole and the
zero-mass solution, and $d^{D-2}x_{\mu\nu}$ represents the
integration over the co-dimension two boundary. For the examples
treated in this paper, the action can schematically be written as
$$
S=\int d^Dx\,\sqrt{-g}\,{\cal L}(g, \phi, A_{(i)}),
$$
where $\phi$ is a scalar field (possibly a dilatonic field) with its
usual kinetic term and $A_{(i)}=A_{(i)\mu}dx^{\mu}$ stand for
Abelian gauge fields or Proca fields. In this generic case, the
boundary term and Noether potential needed to compute the charge
(\ref{quasilocalcharge}) are given by
\begin{eqnarray}
\Theta^\mu &=&2\sqrt{-g}\Big[P^{\mu(\alpha
\beta)\gamma}\nabla_\gamma\delta g_{\alpha\beta} -\delta
g_{\alpha\beta}\nabla_\gamma P^{\mu(\alpha\beta)\gamma}
+\frac{1}{2}\,\displaystyle{\sum_{i}}\left(\frac{\partial
\mathcal{L}}{\partial \left(\partial_{\mu}A_{(i)\nu}\right)}\delta
A_{(i)\nu}\right) +\frac{1}{2}\,\frac{\partial \mathcal{L}}
{\partial \big(\partial_{\mu}\,\phi\big)}\delta \phi\Big],
\nonumber\\
K^{\mu\nu}
&=&\sqrt{-g}\,\left[2P^{\mu\nu\rho\sigma}\nabla_\rho\xi_\sigma
-4\xi_\sigma\nabla_\rho P^{\mu\nu\rho\sigma}\right.
\left.-\displaystyle{\sum_{i}}\,\frac{\partial \mathcal{L}}{\partial
\left(\partial_{\mu}A_{(i)\nu}\right)} \xi^{\sigma}
A_{(i)\sigma}\right] \label{eq:K},
\end{eqnarray}
where $P^{\mu\nu\rho\sigma}=\frac{\partial {\cal L}}{\partial
R_{\mu\nu\rho\sigma}}$, and $R_{\mu\nu\rho\sigma}$ is the Riemann
tensor.

The plan of the paper is organized as follows. In the next section,
a general formula for the asymptotic growth of the number of states
including the angular momentum is proposed. This generic Cardy-like
formula involves the effective spatial dimension
$d_{\tiny{\mbox{eff}}}$, the Lifshitz dynamical exponent $z$, the
mass and angular momentum of the black hole as well as the vacuum
energy which corresponds to the mass of the bulk soliton. In Sec.
$3$, we corroborate the validity of the Cardy-like formula in the
isotropic case $z=1$ with stationary cylindrical black holes. The
case of a three-dimensional black hole solution of the Einstein
equations with a source given by a self-interacting scalar field
with a super-renormalizable potential is also treated in full
details. Lovelock AdS black holes will also be inspected in order to
reinforce the validity of the Cardy-like formula. To end the Sec.
$3$, two examples of hyperscaling violating black holes with
different effective spatial dimensionality will be studied. In Sec.
$4$, we will deal with the anisotropic case $z\not=1$. Lifshitz
black holes solutions of higher-order gravity theories will be our
first testing example while the case of charged anisotropic black
holes produced by various dilaton fields will constitute our second
class of example. Finally, the last section is devoted to the
summary and to the concluding remarks. For simplicity, we have
decided to fix the radius of curvature to unit, $l=1$, while the
Newton gravitational constant $G$ is defined through the change
$2\kappa=16\pi G$.

\section{General formula for the asymptotic growth of the number of states}
As recalled in the introduction, the asymptotic symmetries of
AdS$_3$ are represented by two copies of the Virasoro algebra with
equal left and right moving central charges
$$
c^{+}=c^{-}=c=\frac{3l}{2G}=\frac{12\pi}{\kappa},
$$
(in our convention $l=1$ and $2\kappa=16\pi G$), and the standard
Cardy formula takes the following form
\begin{eqnarray}
S=2\pi \sqrt{\frac{c}{6}\,\widetilde{\Delta}^{+}}+2\pi
\sqrt{\frac{c}{6}\,\widetilde{\Delta}^{-}},\label{Cardystandard}
\end{eqnarray}
where $\widetilde{\Delta}^{\pm}=\frac{1}{2}(M\pm J)$ are the
eigenvalues of the left and right Virasoro operators. In this
representation of the Cardy formula, it is implicitly assumed that
the ground state is identified with the AdS spacetime. Nevertheless,
the AdS spacetime is only a suitable ground state in the case of
standard General Relativity, and this assumption is not longer valid
in presence of source for example. Hence, it is more reasonable to
deal with a Cardy formula involving the vacuum charge than the
central charge. In the vacuum sector for standard General
Relativity, the ground state is nothing but the three-dimensional
AdS soliton whose mass is computed below (\ref{solitonD}) and gives
$M_{\tiny{\mbox{sol}}}=-\pi/\kappa$. Finally, the standard Cardy
formula (\ref{Cardystandard}) can be as well expressed as
\begin{eqnarray}
S=4\pi\sqrt{-\frac{1}{2}M_{\tiny{\mbox{sol}}}^{\tiny{\mbox{}}}}
\sqrt{\widetilde{\Delta}^{-}}+4\pi\sqrt{-\frac{1}{2}M_{\tiny{\mbox{sol}}}^{\tiny{\mbox{}}}}
\sqrt{\widetilde{\Delta}^{+}}. \label{Cardy3d}
\end{eqnarray}
Notice that Cardy-like formulas involving the vacuum energy instead
of the central charges have been proved to be very useful for
examples where the ground state is not the three-dimensional AdS
spacetime, see e. g.
\cite{Gonzalez:2011nz,Shaghoulian:2015dwa,Bravo-Gaete:2015wua,Correa:2010hf,Correa:2011dt,Correa:2012rc}.
The matching between the gravitational entropy and the Cardy formula
(\ref{Cardy3d}) is perfectly consolidated for three-dimensional
black holes that are asymptotically AdS (even in the weaker sense).
Nevertheless, as mentioned in the introduction, we are interested on
generalizations of the Cardy formula that apply for field theories
displaying an hyperscaling violation behavior such that the thermal
entropy ${\cal S}$ scales w. r. t. the temperature $T$ as
$$
{\cal S}\sim T^{\frac{d_{\tiny{\mbox{eff}}}}{z}},
$$
where $d_{\tiny{\mbox{eff}}}$ is an effective spatial dimensionality
and $z$ is the Lifshitz exponent. In order to achieve this task, we
closely follow the derivations done in Refs.
\cite{Shaghoulian:2015dwa, Shaghoulian:2015lcn}. The partition
function ${\cal Z}$ defined on the torus of modulus $\tau$ such that
$ 2\pi\tau =2\pi r\,e^{i\phi}$ and $ 2\pi\bar{\tau} =2\pi
r\,e^{-i\phi}$ can be written as
$$
{\cal Z}[\tau,\bar{\tau}]=\mbox{Tr}\Big[e^{2\pi i\tau L_0}e^{-2\pi
i\bar{\tau} \bar{L}_0}\Big],
$$
with $L_0+\bar{L}_0=M$ and $L_0-\bar{L}_0=J$. The density of states
$\rho(M,J)$ can be obtained by taking an inverse Laplace transform
yielding
\begin{eqnarray}
\rho(M,J)=\int dr\,d\phi\, {\cal Z}[r,\phi]\exp\Big[-2\pi i
re^{i\phi}L_0+2\pi i re^{-i\phi}\bar{L}_0\Big].
\end{eqnarray}
In the microcanonical ensemble, the entropy is basically the
logarithm of the density of states ${\cal S}\sim \log\rho(M,J)$.
Defining the quantity
$$
{\cal Z}_0[r,\phi]=\mbox{Tr}\Big\{\exp\left[2\pi i r
e^{i\phi}\left(L_0-\frac{M_{\tiny{\mbox{sol}}}}{2}\right)- 2\pi i r
e^{-i\phi}\left(\bar{L}_0-\frac{M_{\tiny{\mbox{sol}}}}{2}\right)\right]\Big\},
$$
and assuming that ${\cal Z}_0$ presents the following modular
invariance
$$
{\cal
Z}_0\Big[-\frac{1}{r^{\frac{d_{\mbox{\tiny{eff}}}}{z}}},-\phi\Big]={\cal
Z}_0[r,\phi],
$$
the density of states $\rho(M,J)$ can be re-written as
$$
\rho(M,J)=\int dr\,d\phi\, {\cal
Z}_0\Big[-\frac{1}{r^{\frac{d_{\mbox{\tiny{eff}}}}{z}}},-\phi\Big]\,
\exp\Big[-\frac{\pi i
M_{\tiny{\mbox{sol}}}}{r^{\frac{d_{\mbox{\tiny{eff}}}}{z}}}e^{-i\phi}+
\frac{\pi i
M_{\tiny{\mbox{sol}}}}{r^{\frac{d_{\mbox{\tiny{eff}}}}{z}}}e^{i\phi}-2\pi
i r e^{i\phi}L_0+2\pi i r e^{-i\phi}\bar{L}_0\Big].
$$
Now, as usual, this last expression can be evaluated using a
saddle-point approximations for $r$ and $\phi$, and assuming that
${\cal Z}_0$ varies slowly, one gets
\begin{eqnarray}
S=&&\pi
\sqrt{\frac{d_{\mbox{\tiny{eff}}}+z}{z}}\Big[\left(-2M_{\tiny{\mbox{sol}}}\right)^z
\frac{1}{d_{\mbox{\tiny{eff}}}^{d_{\mbox{\tiny{eff}}}}}\Big]^{\frac{1}{z+d_{\mbox{\tiny{eff}}}}}
\left(\sqrt{(d_{\mbox{\tiny{eff}}}+z)^2M^2-4d_{\mbox{\tiny{eff}}} zJ^2}+(d_{\mbox{\tiny{eff}}}+z)M\right)^\frac{1}{2}\nonumber\\
&&\times \left(\sqrt{\left(d_{\mbox{\tiny{eff}}}+z\right)^2M^2-4
d_{\mbox{\tiny{eff}}}zJ^2}-(d_{\mbox{\tiny{eff}}}-z)M\right)^\frac{d_{\mbox{\tiny{eff}}}-z}{2(d_{\mbox{\tiny{eff}}}+z)}.
\label{CardyHvm}
\end{eqnarray}
This formula constitutes the extension of the Cardy formula
(\ref{Cardy3d}) in arbitrary dimension for a field theory with a
spatial effective dimension $d_{\mbox{\tiny{eff}}}$ and dynamical
exponent $z$.

Let us see the consistency of this expression with known formulas.
First of all, in the isotropic case $z=1$, the expression
(\ref{CardyHvm}) is compatible with the Cardy formula
(\ref{Cardy3d}) in the standard AdS case in three dimensions (which
corresponds to $d_{\mbox{\tiny{eff}}}=1$) as well as with the
formula derived in \cite{Shaghoulian:2015lcn} for hyperscaling
violation metric. On the other hand, in the non-rotating case  with
anisotropy, i. e. $J=0$ with $z\not=1$,  the formula
(\ref{CardyHvm}) reproduces the Lifshitz Cardy formula in three
dimensions with $d_{\mbox{\tiny{eff}}}=1$, see
\cite{Gonzalez:2011nz}, and also the generic formula for
hyperscaling violation metric
\cite{Shaghoulian:2015dwa,Bravo-Gaete:2015wua}.

In the electrically charged case, the Cardy-like formula
(\ref{CardyHvm}) becomes
\begin{eqnarray}
\label{CardyHvmelec} S=&&\pi
\sqrt{\frac{d_{\mbox{\tiny{eff}}}+z}{z}}\Big[\left(-2M_{\tiny{\mbox{sol}}}\right)^z
\frac{1}{d_{\mbox{\tiny{eff}}}^{d_{\mbox{\tiny{eff}}}}}\Big]^{\frac{1}{z+d_{\mbox{\tiny{eff}}}}}\nonumber\\
&&\times \left(\sqrt{(d_{\mbox{\tiny{eff}}}+z)^2\left(M-\frac{1}{2}\phi_e Q_e\right)^2-4d_{\mbox{\tiny{eff}}} zJ^2}+(d_{\mbox{\tiny{eff}}}+z)\left(M-\frac{1}{2}\phi_e Q_e\right)\right)^\frac{1}{2}\\
&&\times
\left(\sqrt{\left(d_{\mbox{\tiny{eff}}}+z\right)^2\left(M-\frac{1}{2}\phi_e
Q_e\right)^2-4 d_{\mbox{\tiny{eff}}}z
J^2}-(d_{\mbox{\tiny{eff}}}-z)\left(M-\frac{1}{2}\phi_e
Q_e\right)\right)^\frac{d_{\mbox{\tiny{eff}}}-z}{2(d_{\mbox{\tiny{eff}}}+z)},\nonumber
\end{eqnarray}
where $\phi_e$ denotes the electric potential while $Q_e$ stands for
the electric charge.

\section{Corroborating the Cardy-like formula in the isotropic case, $z=1$}
In this section, we will be mainly concerned with planar black holes
that are asymptotically AdS or exhibiting an hyperscaling violation
behavior. In these cases, the asymptotic form of the metric can be
parameterized as follows
\begin{eqnarray}
ds^2=\frac{1}{r^{\frac{2\theta}{D-2}}}\Big[-r^2dt^2+\frac{dr^2}{r^2}+r^2\sum_{i=1}^{D-2}dx_i^2\Big],
\label{asmet}
\end{eqnarray}
where $\theta$ represents the parameter responsible of the violation
of the hyperscaling property, and $\theta=0$ will correspond to the
planar AdS case. Note that, for the asymptotic metric (\ref{asmet}),
the isotropic transformations $t\to \lambda t$, $r\to \lambda^{-1}
r$ and $x_i\to \lambda x_i$ are identified as an isometry in the AdS
case, and as a conformal transformation for non vanishing $\theta$.

For this class of black holes, the dynamical exponent  appearing in
the Cardy-like formula (\ref{CardyHvm}) corresponds to the isotropic
situation $z=1$ while the effective spatial dimension
$d_{\mbox{\tiny{eff}}}$ is given by $d_{\mbox{\tiny{eff}}}=D-2$ in
the AdS case ($\theta=0$) otherwise it will depend explicitly on the
violating parameter $\theta$. In the hyperscaling case, one of the
difficulty is to correctly identify the functional dependence of the
effective spatial dimension. For example, for Einstein gravity with
scalar field source, the effective spatial dimension is
$d_{\mbox{\tiny{eff}}}=D-2-\theta$, while for higher-order gravity
theories, this dependence may be different as shown below, see also
\cite{Bravo-Gaete:2015wua}.

In what follows, we will treat various isotropic examples with the
aim of testing the Cardy-like formula, starting from AdS planar
black holes. As a first example, we examine $D-$dimensional
stationary cylindrical black holes solutions of Einstein gravity
with a negative cosmological constant. We also look at the case of a
three-dimensional solution of Einstein gravity with a
self-interacting scalar field with a super-renormalizable potential.
The case of higher theories is also inspected through the analysis
of Lovelock AdS black holes. In the second part, two examples of
hyperscaling violation black holes will be studied. The first
testing example is described by the Einstein gravity with a scalar
field source corresponding to a spatial dimensionality
$d_{\mbox{\tiny{eff}}}=D-2-\theta$. Next, in order to test the force
of the Cardy-like formula with a different spatial dimensionality,
we deal with an hyperscaling violating black hole solution of a pure
quadratic gravity  theory.

\subsection{Stationary cylindrical black holes }
We start the corroborating study of the Cardy-like formula
(\ref{CardyHvm}) with the case of the Einstein field equations in
the presence of a negative cosmological constant
$$
G_{\mu\nu}-\frac{(D-1)(D-2)}{2}g_{\mu\nu}=0,
$$
and whose corresponding action is
\begin{eqnarray}
S[g]=\frac{1}{2\kappa}\int d^Dx\,
\sqrt{-g}\,\Bigg(R+{(D-1)(D-2)}\Bigg).
\end{eqnarray}
We consider the higher-dimensional extension of the stationary
cylindrical black hole found by Lemos \cite{Lemos:1994xp} in four
dimensions and reported in \cite{Awad:2002cz},
\begin{eqnarray}
ds^2=&&-F(r)\left(\Xi\,dt-\sum_{i=1}^{n}\,a_i\,d\phi_i\right)^2+{r^2}\sum_{i=1}^{n}({a_i
}\, dt-\Xi \,d\phi_i)^2 +{dr^2 \over
F(r)}\nonumber\\
&&-{r^2 }\sum_{i<j}^n
\left(a_i\,d\phi_j-a_j\,d\phi_i\right)^2+{r^2}\sum_{i=1}^{D-2-n}dx_i^2.
\label{LemosD}
\end{eqnarray}
Here $n=[(D-1)/2]$ corresponds to the number of rotation parameters
$a_i$, \,$\Xi=\sqrt{1+\sum_i^n \,a_i^2}$, and the metric function
reads
$$
F(r)={r^2}\Big(1-\left(\frac{r_h}{r}\right)^{D-1}\Big).
$$
In four dimensions, the number of rotations is $n=1$, and the
solution reduces to the stationary cylindrical black hole solution
of Lemos \cite{Lemos:1994xp}.

As calculated in Ref. \cite{Awad:2002cz}, the entropy of the
solution is
\begin{eqnarray}
{\cal S}=\frac{2\pi\mbox{Vol}(\Sigma_{D-2}) \Xi r_h^{D-2}}{\kappa},
\label{entLD}
\end{eqnarray}
with mass and angular momenta given by
\begin{eqnarray}
M=\frac{\mbox{Vol}(\Sigma_{D-2})}{2\kappa
}\Big((D-1)\Xi^2-1\Big)r_h^{D-1},\qquad
J_i=\frac{(D-1)\mbox{Vol}(\Sigma_{D-2})}{2\kappa }\,\Xi\,a_i\,
r_h^{D-1}, \label{massJiLD}
\end{eqnarray}
where $\mbox{Vol}(\Sigma_{D-2})$ corresponds to the volume element
of the $(D-2)$-dimensional Euclidean space.

In order to test the validity of the Cardy-like formula
(\ref{CardyHvm}), one needs to construct the static soliton and to
compute its mass through the quasilocal expression
(\ref{quasilocalcharge}). As explained in the introduction,
operating a double Wick rotation on the static version of the
solution, that is Eq.  (\ref{LemosD}) with $a_i=0$, one gets the AdS
soliton \cite{Horowitz:1998ha}
\begin{equation}
 d{s}^2 = -r^2\,{{d{t}}}^{2}+
 \frac{d{r}^{2}}{f({r})}+f({r})  \,d{{\varphi}^{2}}+
 r^2\,\sum_{i=1}^{D-3} d{{x}_{i}^{2}},
 \label{solitonLemosD}
\end{equation}
with
\begin{eqnarray}
f({r})={r}^{2}
 \left[1-\left(\frac{2}{(D-1) {r}}\right)^{D-1}\right].
\label{solitonLemosfunctions}
\end{eqnarray}
The next step is to determine the mass of the soliton through the
quasilocal formula (\ref{quasilocalcharge}) where the Killing vector
field is $\xi^{{t}}=\partial_{{t}}$. The variation of the Noether
potential and the surface term are given by
\begin{eqnarray}
\int_{0}^{1}\!\!\!\!ds\,\Theta^{{r}}= \frac{1}{2 \kappa}
\left(\frac{2}{D-1}\right)^{D-1} ,\qquad \Delta K^{{r} {t}}
(\xi^{{t}})=-\frac{1}{\kappa} \left(\frac{2}{D-1}\right)^{D-1}.
\end{eqnarray}
Finally, the mass of the $D-$dimensional  AdS gravitational soliton
reads
\begin{eqnarray}
M_{\tiny{\mbox{sol}}}=-\frac{\mbox{Vol}(\Sigma_{D-2})}{2 \kappa}
\left(\frac{2}{D-1}\right)^{D-1}, \label{solitonD}
\end{eqnarray}
and it is simple to check that the formula (\ref{CardyHvm}) where
$J^2$ is now understood as $J^2=\sum_{i=1}^{n}J_i^2$ and where
$d_{\mbox{\tiny{eff}}}=D-2$ and $z=1$ correctly reproduces the
gravitational entropy (\ref{entLD}).

\subsection{Black hole with a super-renormalizable self-interacting scalar field in $3$D}
We pursue our survey considering now a three-dimensional toy model
whose action is described by the Einstein-Hilbert piece with a
cosmological constant together with a nonminimally self-interacting
scalar field
\begin{align}
S[g,\phi]={}&\int\!\!d^{3}x\sqrt{-g}\left(\frac{R-2\Lambda}{2\kappa}-
\frac{1}{2}\nabla_{\mu}\phi\nabla^{\mu}\phi-\frac{1}{16}R\phi^2-U(\phi)\right).
\label{actionscalarfield}
\end{align}
The nonminimal coupling corresponds to the conformal one in three
dimensions, and as it is well known the potential term which is
compatible with the conformal invariance of the matter source is
$U(\phi)\propto \phi^6$. Nevertheless, in our case, we chose a
potential term breaking the conformal invariance of the matter
action, and defined by all the powers lower than the conformal one
(super-renormalizable potential), that is
\begin{eqnarray}\label{potAHM}
U(\phi)=\lambda_1\,\phi+\lambda_2\,\phi^{2}+
\lambda_3\,\phi^{3}+\lambda_4\,\phi^{4}+\lambda_5\,\phi^{5}
+\lambda_6\,\phi^{6}.
\end{eqnarray}
The field equations obtained from the variation of the action
(\ref{actionscalarfield}) with respect to the metric and the scalar
field are
\begin{eqnarray}
&&G_{\mu\nu}+\Lambda g_{\mu\nu}=\kappa
\left(\nabla_{\mu}\phi\nabla_{\nu}\phi
-g_{\mu\nu}\left(\frac{1}{2}\nabla_{\sigma}\phi\nabla^{\sigma}\phi
+U\right)+\frac{1}{8}( g_{\mu\nu}\Box - \nabla_{\mu}\nabla_{\nu}+G_{\mu\nu} )\phi^2\right),\nonumber\\
&&\Box\phi-\frac{1}{8}R\phi=\frac{dU}{d\phi}. \label{superrenfeqs}
\end{eqnarray}
In Ref. \cite{Ayon-Beato:2015ada}, the authors have derived a static
black hole solution of the model described by
(\ref{actionscalarfield}-\ref{potAHM}-\ref{superrenfeqs}) using a
conformal machinery where the coupling constants are parameterized
as follows
\begin{eqnarray} \label{lambdas}
\lambda_1&=&{\frac { \big[ \left( \mu-3 \right) ^{2} \left( 4\,\mu-3
 \right) \lambda^{4}+27\, \left( \mu-1 \right) ^{2} \big]
 \lambda\sqrt {2}}{
18 \sqrt {\kappa} \left( \mu-1 \right) ^{2}
 \left( 1-\lambda^{2} \right)
^{5 } }} ,\nonumber\\
\lambda_2&=&-{\frac {5 \lambda^{2} \big[ \lambda^{2} \left( 4\,\mu-3
 \right)  \left( \mu-3 \right) ^{2}+27\, \left( \mu-1 \right) ^{2}
 \big] }{ 72 \left( \mu-1 \right) ^{2} \left( 1-\lambda^{2} \right) ^{5}
}},\nonumber\\
\lambda_3&=&{\frac {5 \sqrt { 2\kappa} {\mu}^{3}\lambda^{3}}{
 54 \left( \mu-1 \right) ^{2} \left( 1-\lambda^{2} \right) ^{5}
}},\qquad \lambda_4=-{\frac {5 \kappa\,\lambda^{2} \big[ \left(
4\,\mu-3
 \right)  \left( \mu-3 \right) ^{2}+27\, \left( \mu-1 \right) ^{2}\lambda^
{2} \big] }{576 \left( \mu-1 \right) ^{2} \left( 1-\lambda^{2}
\right) ^{ 5}}},\nonumber\\
\lambda_5&=&{\frac { \big[ 27\, \left( \mu-1 \right) ^{2}\lambda^
{4}+ \left( 4\,\mu-3 \right)  \left( \mu-3 \right) ^{2} \big] \sqrt
{2}\lambda{\kappa}^{3/2}}{ 1152 \left( \mu-1 \right) ^{2} \left(
1-\lambda^{2}\right) ^{5} }},\\
\lambda_6&=&-{\frac {{\kappa}^{2} \big[27\, \left( \mu-1
 \right) ^{2}\lambda^{6}+ \left( 4\,\mu-3 \right)  \left( \mu-3 \right) ^{
2} \big] }{ 13824 \left( \mu-1 \right) ^{2} \left(
1-\lambda^{2} \right) ^{5 }}},\nonumber\\
\Lambda&=&-{{\frac {\big[27\, \left( \mu-1 \right) ^{2}+\lambda^{6}
\left( 4\,\mu-3
 \right)  \left( \mu-3 \right) ^{2}\big]}
 { 27 \left( \mu-1 \right) ^{2}
\left( 1-\lambda^{2} \right) ^{5}}} }.\nonumber
\end{eqnarray}
More precisely, as shown in \cite{Ayon-Beato:2015ada}, the action
defined by (\ref{actionscalarfield}-\ref{potAHM}-\ref{lambdas}) can
be obtained from the conformally invariant action\footnote{There is
a slight abuse of language in the sense that "by conformally
invariant action", we mean that only  the matter source involving
the scalar field is invariant under the conformal transformations
and not the gravity action.} denoted by
$\tilde{S}[\tilde{g},\tilde{\phi}]$ and corresponding to the action
(\ref{actionscalarfield}) with the potential $U\propto
\tilde{\phi}^6$ through a map parameterized by the factor $\lambda$,
and both actions are related as follows
\begin{eqnarray}
S[g,\phi]=(1-\lambda^2)\,\tilde{S}[\tilde{g},\tilde{\phi}].
\label{relsuperrenoconf}
\end{eqnarray}
In fact, the static solution reported in \cite{Ayon-Beato:2015ada}
was constructed using the one-parameter mapping with a seed
configuration given by the solution of the conformally
self-interacting version of the Martinez-Zanelli solution
\cite{Martinez:1996gn} found in \cite{Henneaux:2002wm}. Instead of
writing down the static solution \cite{Ayon-Beato:2015ada}, we
report its spinning extension obtained from the static configuration
as usual in three dimensions through a Lorentz boost defined by
\begin{equation} t\to
\frac{1}{\sqrt{1-{\omega}^2}}(t+{\omega}\,\varphi),\qquad \varphi\to
\frac{1}{\sqrt{1-{\omega}^2}}(\varphi+{\omega} t),\label{boost}
\end{equation}
and well-defined for $\omega^2<1$. The line element of the resulting
rotating solution is given by
\begin{eqnarray}
ds^2=H^2(r)\left\{-N^2(r)F(r)dt^2+\frac{dr^2}{F(r)}+R^2(r)\left(d\varphi+N^{\varphi}(r)dt\right)^2\right\},
 \label{rotmetric}
\end{eqnarray}
where the metric functions and the scalar field read
\begin{eqnarray}
\label{metfunctions} N^2(r)&=&r^2{\frac {\left( 1-{\omega}^{2}
\right)} { \left( r^2-{\omega}^{2} F(r) \right) }}\, ,\quad
R^2(r)=\frac{1}{(1-{\omega}^{2})} \left( r^2-{\omega}^{2} F(r)
\right),\quad N^{\varphi}(r)={\frac { {\omega} \big( r^2-F(r) \big)
} { \big(
r^2-{\omega}^{2}F(r) \big) }},\nonumber\\
\phi(r)&=&\sqrt {\frac{2}{\kappa\,H(r)}} \left[ \sqrt{{\frac { 12
\left( \mu-1 \right) r_h}{\big(3\, \left( \mu-1
 \right) r_h-2\,r\mu\big)}}
}+2\,\lambda \right]
,\\
F(r)&=&r^2\Bigg[
1+(\mu-1)\left(\frac{r_h}{r}\right)^{3}-\mu\left(\frac{r_h}{r}\right)^{2}\Bigg],\quad
H(r)=\left[ \lambda \,\sqrt{{\frac {3 \left( \mu-1 \right)
r_h}{\big( 3\, \left( \mu-1 \right) r_h-2\,r\mu
 \big) }}}+1 \right] ^{2}.\nonumber
\end{eqnarray}

We now analyze the thermodynamical properties of the spinning
solution through the Euclidean method where the Euclidean time
$\tau$ is imaginary $\tau=it$ and periodic of period $\beta$ which
is the inverse of the temperature $\beta=T^{-1}$. The Euclidean
action $I_{\tiny{{\mbox{Euc}}}}$ is related with the free energy $F$
by
\begin{eqnarray}
I_{\tiny{{\mbox{Euc}}}}=\beta\,F=\beta\left(M-T{\cal S}-\Omega
J\right), \label{freeNRJ}
\end{eqnarray}
where $M$ is the mass, ${\cal S}$ the entropy and $\Omega$ is the
chemical potential corresponding to the angular momentum $J$. On the
other hand, in order to display the boundary term $B$ that will
ensure the finiteness of the Euclidean action, we find more
convenient to consider the following class of Euclidean metric
$$
ds^2=H^2(r)\left\{N^2(r)F(r)d\tau^2+\frac{dr^2}{F(r)}+R^2(r)\left(d\varphi+i
N^{\varphi}(r)d\tau\right)^2\right\},
$$
with the assumption that the scalar field only depends on the radial
coordinate, $\phi=\phi(r)$. The Euclidean time $\tau\in [0,\beta]$
and the radial coordinate $r\in [r_h,\infty[$ where $r_h$ is the
location of the horizon and $\varphi\in [0,2\pi[$. The reduced
action principle reads
\begin{eqnarray}
\label{redaction} I_{\tiny{{\mbox{Euc}}}}= 2\pi\beta \int
\left(N(r)\mathcal{H}(r) + {N^{\varphi}(r)} p(r)^{\prime} \right)dr
+B,
\end{eqnarray}
where
\begin{eqnarray}
p(r)= \frac{1}{16}\frac{H(r) R^3(r)(8 -
\phi(r)^2\kappa){N^{\varphi}(r)}^{\prime}}{N(r)\kappa}, \label{eqp}
\end{eqnarray}
and the Hamiltonian $\mathcal{H}$ is given by
\begin{eqnarray*}
\mathcal{H} = && \dfrac{8-\kappa \phi^2}{8\kappa} \left[ RFH^{''} +  H FR^{''} - \dfrac{RF H^{'2}}{H} + \dfrac{1}{2}F^{'}R^{'}H + H^{'}\left(\dfrac{1}{2}F^{'}R + R^{'}F \right)  \right] \\ \nonumber && - \dfrac{1}{4}H FR\phi \phi^{''} + \dfrac{1}{4}H FR\phi^{'2} - \dfrac{1}{4}\left(\dfrac{1}{2}F^{'}R + FR^{'} \right)H\phi\phi^{'} + \dfrac{H^3 R(\Lambda + \kappa U(\phi))}{\kappa} \\ \nonumber && - \dfrac{24\kappa \,p^2}{H R^3(\kappa \phi^2 - 8)}.\\
\end{eqnarray*}
In the reduced action (\ref{redaction}), $B$ is a boundary term that
is fixed by requiring that the Euclidean action has an extremum,
that is $\delta I_{E}=0$ ; this last condition in turn implies that
\begin{eqnarray*}
&& \delta B=  -2\pi\beta \left[ \left(
\dfrac{8-\kappa\phi^2}{8\kappa} \left\{ \dfrac{1}{2} H^{'}RN +
\dfrac{1}{2}H R^{'}N \right\} - \dfrac{1}{8}H RN\phi \phi^{'}
\right) \delta F + \left( \dfrac{8-\kappa \phi^2}{8\kappa}RFN
\right) \delta H^{'} + \right.
\\ \nonumber
&& \left. + \left( \dfrac{8-\kappa \phi^2}{8\kappa} \left\{
-2\dfrac{H^{'}RFN}{H} - \dfrac{1}{2}RNF^{'} - RFN^{'} \right\} +
\dfrac{1}{4}RFN\phi \phi^{'} \right) \delta H + \left(
\dfrac{8-\kappa \phi^2}{8\kappa} H FN \right) \delta R^{'} \right.
 \\ \nonumber
&& \left. - \left( \dfrac{8-\kappa \phi^2}{8\kappa} H \left\{
\dfrac{1}{2}F^{'}N + FN^{'} \right\}  \right) \delta R - \left(
\dfrac{1}{4}H FRN\phi \right) \delta \phi^{'} + \left( \dfrac{3}{4}
H FRN\phi^{'} + \dfrac{1}{8}H F^{'}RN\phi  \right. \right.
 \\ \nonumber
&& \left. \left. +\ \dfrac{1}{4}H FRN^{'}\phi +
\dfrac{1}{4}H^{'}FRN\phi \right) \delta \phi  + N^{\varphi}\delta p
\right]_{r=r_h}^{r=\infty},
\end{eqnarray*}
where the variation is taken between the horizon and the infinity.
The temperature is fixed requiring regularity of the metric at the
horizon yielding to
\begin{equation*}
\beta (N(r)F^{'}(r))|_{r_h} = 4\pi,
\end{equation*}
and for the solution (\ref{metfunctions}), one obtains
\begin{equation}
T = \dfrac{(3-\mu)r_h \sqrt{1-\omega^2}}{4\pi }.
\end{equation}
We do not display the field equations of the reduced action
(\ref{redaction}) but their full integration will reproduce the
solution (\ref{metfunctions}) with
$$
p=- \dfrac{\omega \mu (1-\lambda^2) }{\kappa  (1-\omega^2)} r_h^2.
$$
We are now in position to compute the boundary term. Its
contribution at the infinity gives
\begin{eqnarray*} \delta B\big|_{\infty} = \dfrac{2\pi \beta \mu (1+ \omega^2)(1-\lambda^2) r_h}{ \kappa (1-\omega^2)} {\delta r_h}\Longrightarrow
B\big|_{\infty} = \beta \dfrac{\pi \mu
(1+\omega^2)(1-\lambda^2)}{\kappa (1-\omega^2)} r_h^2,
\end{eqnarray*}
while at the horizon, one gets
\begin{eqnarray*}
\delta B\big|_{r_{h}} = -2\pi\beta \left[
\dfrac{4\pi(1-\lambda^2)\mu}{\sqrt{1-\omega^2}(\mu -3) \kappa \beta}
\delta r_{h} + N^{\varphi}(r_h)\ \delta p \right],
\end{eqnarray*}
and since $\Omega = N^{\varphi}(\infty) - N^{\varphi}(r_h) = -
\omega $, we obtain
\begin{eqnarray}
B\big|_{r_{h}} = \dfrac{8\pi^2(1-\lambda^2)\mu}{\sqrt{1-\omega^2}(3-
\mu) \kappa} r_{h} + 2\pi \beta \Omega p.
\end{eqnarray}
Finally, the boundary term is given by
\begin{eqnarray}
\nonumber  B && = B\big|_{\infty} - B\big|_{r_{h}}\\  && = \beta
\dfrac{\pi \mu (1+\omega^2)(1-\lambda^2)}{\kappa (1-\omega^2)} r_h^2
- \dfrac{8\pi^2(1-\lambda^2)\mu}{\sqrt{1-\omega^2}(3- \mu) \kappa}
r_{h} - 2\pi \beta \Omega p, \label{BT}
\end{eqnarray}
and hence the comparison  between (\ref{BT}) and (\ref{freeNRJ})
permits the identification of the entropy
\begin{eqnarray}
&&{ \cal S}={\frac {8 {\mu} {\pi^2}{r_h}  \left( 1-{\lambda}^{2}
\right) }{\kappa \, \left( 3-\mu \right) \sqrt{1-{\omega}^{2}}}},
\label{entropy3d}
\end{eqnarray}
as well as the mass, angular momentum  and angular velocity that are
given by
\begin{eqnarray}
\label{thermoqtes} && M={ \frac {\mu \pi
(1+{\omega}^{2})(1-\lambda^{2}) }{
\kappa\,(1-{\omega}^{2})}}r_h^2,\quad J=-{ \frac { 2\pi\mu {\omega}
(1-\lambda^{2}) }{\kappa\,(1-{\omega}^{2})}} r_h^{2},\quad
\Omega=-\omega.
\end{eqnarray}
It is interesting to note that the values of the entropy, the mass
and the angular momentum of the black hole solution with the
super-renormalizable potential are precisely $(1-\lambda^2)$ time
those of the black hole solution with the conformal potential
\cite{Martinez:1996gn, Henneaux:2002wm}. This is not surprising and
can be explained as follows. On one hand, the actions being
proportional with that precise factor (\ref{relsuperrenoconf}), so
that their Euclidean actions
$$
I_{\tiny{{\mbox{Euc}}}}=(1-\lambda^2)\tilde{I}_{\tiny{{\mbox{Euc}}}}.
$$
On the other hand, since the temperatures and the chemical
potentials are the same for both solution, $T=\tilde{T}$ and
$\Omega=\tilde{\Omega}$, we have by virtue of (\ref{freeNRJ})
$$
I_{\tiny{{\mbox{Euc}}}}=\beta\left(M-T{\cal S}-\Omega
J\right)=(1-\lambda^2)\tilde{I}_{\tiny{{\mbox{Euc}}}}=(1-\lambda^2)\beta\left(\tilde{M}-T{\cal
\tilde{S}}-{\Omega} \tilde{J}\right).
$$
For completeness, we also notice that the first law of
thermodynamics is satisfied
\begin{eqnarray}
dM=Td{\cal S}-\omega\, dJ.\label{firstlaw}
\end{eqnarray}
Once again, in order  to display the role played by the
gravitational soliton for computing the entropy, we construct the
gravitational soliton and derive its mass. The soliton obtained from
the static black hole configuration, Eqs
(\ref{rotmetric}-\ref{metfunctions}) with $\omega=0$, through a
double Wick rotation reads
\begin{equation}
 d{s}^2 = -r^2\,h^2({r}){{d{t}}}^{2}+
 \frac{h^2({r})\,d{r}^{2}}{{r}^2f({r})}+{{{r}}^{2} f({r})}\,h^2({r}) d{\varphi}^{2},
 \label{metricsoliton}
\end{equation}
with
\begin{eqnarray*}
f({r})= 1+(\mu-1)\left[\frac {2 }{\left(3- \mu \right) {r}
}\right]^{3} -\mu \left[\frac {2 }{\left(3- \mu \right) {r}
}\right]^{2},\quad h({r})=\left[ \lambda \,\sqrt{{\frac {3  \left(
\mu-1 \right) } {\big( 3  \left( \mu-1 \right)-{r}\mu
\left(3-\mu\right)\big) }}}+1 \right] ^{2}.
\end{eqnarray*}
For a Killing vector $\xi^{{t}}=(1,0,0)$, the surface term and the
variation of the Noether potential read
\begin{eqnarray*}
\int_{0}^{1}\!\!\!\!ds\,\Theta^{{r}}&=& (1-\lambda^{2})\,\left[-{
\frac { 8{\mu}^{2} {r}}{9 \left( 3-\mu \right) \kappa\, \left( \mu-1
\right) }}+{\frac {{\mu}^{2} \left( 4\,\mu-3 \right)  \left( \mu-3
\right) ^{4}{{r}}^{4}}{27  \left( \mu-1 \right) ^{2} \big( 3  \left(
\mu-1 \right) +r\mu\, \left( \mu-3 \right) \big) ^{2}
\kappa}}\right.\nonumber\\
&-&\left.{\frac { \left( 4\,\mu-3 \right) \left( \mu-3 \right)
^{2}{{r}}^{2}}{27 \left( \mu-1 \right) ^{2}\kappa}} -{\frac
{2 \mu}{ \left( 3-\mu \right) ^{2}\kappa}}\right],\nonumber\\
\Delta K^{{r} {t}} (\xi^{{t}})&=&(1-\lambda^{2})\,\left[{ \frac {
8{\mu}^{2} {r}}{9 \left( 3-\mu \right) \kappa\, \left( \mu-1 \right)
}}-{\frac {{\mu}^{2} \left( 4\,\mu-3 \right)  \left( \mu-3 \right)
^{4}{{r}}^{4}}{27  \left( \mu-1 \right) ^{2} \big( 3 \left( \mu-1
\right) +r\mu\, \left( \mu-3 \right) \big) ^{2}
\kappa}}\right.\nonumber\\
&+&\left.{\frac { \left( 4\,\mu-3 \right) \left( \mu-3 \right)
^{2}{{r}}^{2}}{27 \left( \mu-1 \right) ^{2}\kappa}} \right],
\end{eqnarray*}
yielding to a  mass of the soliton (\ref{quasilocalcharge}) given by
\begin{eqnarray}
M_{\mbox{\tiny{sol}}}=-\frac{4\pi\mu
(1-\lambda^2)}{(3-\mu)^2\kappa}. \label{masssolAHM}
\end{eqnarray}
Finally, it is easy to check that the formula of the gravitational
entropy (\ref{entropy3d}) is correctly reproduced by means of the
Cardy-like formula (\ref{CardyHvm}) using the conserved quantities
(\ref{thermoqtes}) and the mass of the gravitational soliton
(\ref{masssolAHM}).

\subsection{Lovelock AdS black holes}
In General Relativity, two of the main fundamental assumptions are
the requirement of general covariance and the fact that the field
equations for the metric are at most of second order. In three and
four dimensions, these requirements automatically single out the
gravity theory to be described by the Einstein-Hilbert action plus
eventually a cosmological constant. However, for dimensions greater
than four, a more general gravity theory, the so-called Lovelock
theory, satisfies these standard requirements \cite{LOV}. The
$D-$dimensional Lovelock Lagrangian is a $D-$form constructed out of
the vielbein, the spin connection and their exterior derivative and
is given by
\begin{eqnarray*}
&& \sum_{p=0}^{[D/2]}\alpha_p\,\,\epsilon_{a_1\cdots a_D} R^{a_1a_2}
\cdots
 R^{a_{2p-1}a_{2p}} e^{a_{2p+1}}\cdots   e^{a_D},
\end{eqnarray*}
where $R^{ab} = d\,\omega^{ab} + \omega^{a}_{\;c}\,\omega^{cb}$ is
the curvature two-form, the coefficients $\alpha_p$ are arbitrary
dimensionful coupling constants and the wedge product between
differential forms is understood. We recognize the first two terms
of the Lovelock Lagrangian to be proportional to the cosmological
constant and to the Einstein-Hilbert piece. Being $D-$dimensional
forms, the Lovelock actions are automatically invariant under the
local Lorentz transformations. In addition, in odd dimension, this
Lorentz gauge symmetry can be enlarged for a particular choice of
the coefficients $\alpha_p$ to a local (A)dS or Poincar\'e symmetry
group ; the resulting Lagrangians are  called Chern-Simons, see e.
g.  \cite{CSsugra} for a review on Chern-Simons theory. As shown in
\cite{Crisostomo:2000bb}, the coefficients $\alpha_p$ can also be
chosen such that the theory has a unique AdS vacuum with a fixed
value of the cosmological constant. In doing so, one yields to a
series of inequivalent actions indexed by an integer $n$ with $1\leq
n\leq [(D-1)/2]$, and given by
\begin{eqnarray}
S_n&=&-\frac{1}{2\kappa n(D-3)!}\int
\sum_{p=0}^{n}\frac{C^n_p}{(D-2p)}\epsilon_{a_1\cdots a_D}
R^{a_1a_2} \cdots
 R^{a_{2p-1}a_{2p}} e^{a_{2p+1}}\cdots  e^{a_D},
 \label{Sk2}
\end{eqnarray}
or in tensorial form by
\begin{eqnarray*}
S_n&=&\frac{1}{2\,\kappa}\int d^{D}x\,\sqrt{-g}
\left[R+\frac{(D-1)(D-2)}{n}
+\frac{(n-1) }{2(D-3)(D-4)}{L}_{GB}\right.\nonumber\\
&+&\left. \frac{(n-1)(n-2)}{3!(D-3)(D-4)(D-5)(D-6)}{L}^{(3)}+\cdots
\right], \label{LOvk}
\end{eqnarray*}
where $L_{\rm
GB}=R^2-4R_{\alpha\beta}R^{\alpha\beta}+R_{\mu\nu\alpha\beta}R^{\mu\nu\alpha\beta}$
stands for the Gauss-Bonnet Lagrangian and ${L}^{(3)}$ is given by
\begin{eqnarray*}
L^{(3)}&=&R^3   -12RR_{\mu \nu } R^{\mu \nu } + 16\,R_{\mu \nu
}R^{\mu }_{\phantom{\mu} \rho }R^{\nu \rho } + 24 R_{\mu \nu
}R_{\rho \sigma }R^{\mu \rho \nu \sigma }+ 3RR_{\mu \nu \rho \sigma
} R^{\mu \nu \rho \sigma }
\nonumber \\
&-&24R_{\mu \nu }R^{\mu} _{\phantom{\mu} \rho \sigma \kappa } R^{\nu
\rho \sigma \kappa  }+ 4 R_{\mu \nu \rho \sigma }R^{\mu \nu \eta
\zeta } R^{\rho \sigma }_{\phantom{\rho \sigma} \eta \zeta }-8R_{\mu
\rho \nu \sigma } R^{\mu \phantom{\eta} \nu \phantom{\zeta}
}_{\phantom{\mu} \eta \phantom{\nu} \zeta } R^{\rho \eta \sigma
\zeta }.\label{cubiclagrangian}
\end{eqnarray*}
Using differential forms, the field equations arising from the
variation of the action  (\ref{Sk2}) with respect to the vielbein
and the spin connection read
\begin{subequations}
\label{Lovkeqs}
\begin{eqnarray}
&&\epsilon_{aa_2\cdots a_{D}}\bar{R}^{a_2a_3}\cdots \bar{R}^{a_{2n-1}a_{2n}}e^{a_{2n+1}}\cdots e^{a_{D}}=0,\\
&&\epsilon_{aba_3\cdots a_{D}}\bar{R}^{a_3a_4}\cdots
\bar{R}^{a_{2n-1}a_{2n}}T^{2n+1}e^{a_{2n+2}}\cdots e^{a_{D}}=0,
\end{eqnarray}
\end{subequations}
where $\bar{R}^{ab}=R^{ab}+e^ae^b$ and $T^a$ is the torsion $2-$form
$T^a=de^a+\omega^a_be^b$. The spectrum of solutions of Lovelock and
Chern-Simons gravity theories contain (topological) AdS black holes
with interesting thermodynamical properties, see e. g.
\cite{Boulware:1985wk,Banados:1993ur,Cai:1998vy,Crisostomo:2000bb,Aros:2000ij,Cai:2001dz}.

We now construct the spinning extension of the black hole solution
of the field equations (\ref{Lovkeqs}) found in \cite{Aros:2000ij}
with planar base manifold. Its line element is given by
\begin{eqnarray}
 ds^2 &=&-N^2(r)
 dt^{2}+\frac{dr^{2}}{F(r)}+R^{2}(r)\,(d\varphi+N^{\varphi}(r)dt)^{2} +r^{2}
 \sum_{i=1}^{D-3} dx_{i}^{2},
\end{eqnarray}
with
\begin{eqnarray}
N^2(r)&=& r ^{2} F (r) \left( 1-{\omega}^{2} \right) \left(r^2-F(r)
{\omega}^{2}\right)^{-1}, \qquad N^{\varphi}(r)=\omega
 \left(r^2-F(r) \right)\,
 \left(r^2-F(r) {\omega}^{2}\right)^{-1},\nonumber\\
R^2(r)&=&\frac{1}{(1-{\omega}^{2})}\,
 \left(r^2-F(r) {\omega}^{2}\right)
, \qquad F(r)=r^{2}
\left(1-\left(\frac{r_{h}}{r}\right)^{\frac{D-1}{n}}\right).
\end{eqnarray}
Skipping the details, the entropy together with the Hawking
temperature read
\begin{eqnarray}
\label{Walda-t-z} \mathcal{S}&=&\frac{2 \pi r_h^{D-2}\,
\mbox{Vol}(\Sigma_{D-2})}{\kappa \sqrt{1-\omega^{2}}},\qquad
T=\frac{(D-1) r_h \sqrt{1-\omega^{2}}}{4 \pi  n },
\end{eqnarray}
while the mass and angular momentum are given by
\begin{equation}\label{mass mom a-t-z}
M=\left(\frac{D-2+\omega^{2}}{2 \kappa}\right)
\,\frac{\mbox{Vol}(\Sigma_{D-2})r_{h}^{D-1}}{n (1-\omega^{2})},
\qquad J=-\left(\frac{D-1}{2 \kappa}\right) \,\frac{\omega
\,\mbox{Vol}(\Sigma_{D-2})r_{h}^{D-1}}{n (1-\omega^{2})},
\end{equation}
and we easily check that the first law  holds.

On the other hand, the corresponding soliton derived from the static
black hole solution with a double analytic continuation is
\begin{eqnarray*}
d{s}^2 &=& -{r}^{2} d {t}^2+ \frac{1}{{r}^2}\,
\frac{d{r}^2}{f({r})}+ {r}^{2}\,f({r})\,d
{\varphi}^{2}+{{r}^{2}}\sum_{i=1}^{D-3}\,d{x}_{i}^{2},\quad
f({r})=1-\left[\frac{ 2 n}{(D-1){r}}\right]^{\frac{D-1}{n}},
\end{eqnarray*}
and the mass of the soliton is computed to be
\begin{eqnarray}\label{massola-t-z}
M_{\mbox{\tiny{sol}}}=-\frac{\mbox{Vol}(\Sigma_{D-2})}{ \kappa
(D-1)}\left(\frac{2 n}{D-1}\right)^{D-2}.
\end{eqnarray}
Finally, the gravitational expression of the entropy
(\ref{Walda-t-z}) matches perfectly with the Cardy-like formula
(\ref{CardyHvm}) with $d_{\mbox{\tiny{eff}}}=D-2$ and with the
conserved quantities (\ref{mass mom a-t-z}-\ref{massola-t-z}).
Notice that this matching is far from trivial and deserves a certain
attention for the following reason. As said in the introduction,
there exists higher-dimensional extension of the Cardy formula that
applied for field theory having an AdS dual, the so-called
Cardy-Verlinde formula \cite{Verlinde:2000wg}. Nevertheless, as
stressed in \cite{Cai:2001jc}, the Cardy-Verlinde formula fails in
general for the Lovelock AdS black holes independently of the
topology of the base manifold.

\subsection{Hyperscaling violation black hole in $D$ dimensions}
Up to now, we have only considered cases where the violating
exponent is vanishing, $\theta=0$. Nevertheless, hyperscaling
violation black holes are also known in the current literature, see
e. g. \cite{hvBH}. It is conjectured that these solutions may have a
certain interest in holographic contexts related to condensed matter
physics, see e. g. \cite{Dong:2012se} and \cite{Alishahiha:2012cm}.
For example, solutions with an hyperscaling violation exponent
$\theta=D-3$ can be useful to describe a dual theory with an ${\cal
O}(N^2)$ Fermi surface ($N$ being the number of degrees of freedom).

Hyperscaling violation black holes can also be an excellent set-up
to test the robustness of the Cardy-like formula (\ref{CardyHvm})
since in this case the effective spatial dimension
$d_{\mbox{\tiny{eff}}}$ is not longer equal to $D-2$ but will
instead depend on the exponent $\theta$. A toy model in order to
achieve this task is given by the Einstein-Hilbert action with a
self-interacting scalar field
\begin{align}
S[g,\phi]={}\int\!\!d^{D}x\sqrt{-g}\Biggl[\frac{R}{2\kappa}-
\frac{1}{2}\nabla_{\mu}\phi\nabla^{\mu}\phi-U(\phi) \Biggr],
\label{actionscalarfieldhvm}
\end{align}
whose field equations read
\begin{eqnarray}
&&G_{\mu\nu}=\kappa \left[\nabla_{\mu}\phi\nabla_{\nu}\phi
-g_{\mu\nu}\left(\frac{1}{2}\nabla_{\sigma}\phi\nabla^{\sigma}\phi
+U\right)\right],\qquad \Box\phi=\frac{dU}{d\phi}. \label{hvmfeqs}
\end{eqnarray}
Indeed, for a Liouville potential of the form
\begin{eqnarray}
U(\phi)&=&-\frac{\left( D-2-\theta \right)  \left(
D-1-\theta\right)}{2\, \kappa} \,{e}^{\frac{ \sqrt{4 \kappa
\theta}\,\phi}{\sqrt{(\theta-D+2)(D-2)}}},\label{potentialES}
\end{eqnarray}
a static hyperscaling violation black hole was found in
\cite{Perlmutter:2012he} with a generic value of the exponent
$\theta$, and whose effective spatial dimension is
$d_{\mbox{\tiny{eff}}}=D-2-\theta$.

As done previously, we construct the spinning extension of the
solution \cite{Perlmutter:2012he} that reads
\begin{eqnarray}
 ds^2 =\frac{1}{r^{\frac{2 \theta}{D-2}}}\left[-N^2(r)
 dt^{2}+\frac{dr^{2}}{F(r)}+R^{2}(r)\,\left(d {\varphi}+N^{\varphi}(r)dt\right)^{2}+{r^{2}} \sum_{i=1}^{D-3} d x_{i}^{2}\right],\label{rotmetricdhvm}
\end{eqnarray}
where
\begin{eqnarray*}\label{hvmrotatingESs}
N^2(r)&=& r^{2}
 {\left( 1-{\omega}^{2}
 \right)}\, F (r)
 \left( r^2-F(r) {\omega}^{2}\right) ^{-1}
, \quad R^2(r)=\frac{1}{1-{\omega}^{2}}\,\left(r^2- F
 \left( r \right) {\omega}^{2}\right)
,\\
N^{\varphi}(r)&=&\omega
\frac{\left(r^2-F(r)\right)}{\left(r^2-\omega^2 F(r)\right)},\quad
F(r)=r^2\Big(1-\left(\frac{r_h}{r}\right)^{D-1-\theta}\Big), \quad
\phi(r)=\sqrt {{\frac {\theta\, \left( \theta-D+2 \right) }{\kappa
(D-2)}}}\ln(r).
\end{eqnarray*}
Since the thermodynamics analysis is quite similar to the case of
the scalar field with a super-renormalizable potential, we only
sketch briefly the quantities of interest as the entropy and
temperature of the solution
\begin{eqnarray}
\mathcal{S}&=& \frac{2 \,\pi\mbox{Vol}(\Sigma_{D-2})}{\kappa
\sqrt{1-{\omega}^{2}}} r_h^{D-2-\theta} ,\qquad T= \frac{r_h
(D-1-\theta)\sqrt{1-{\omega}^{2}}}{4 \pi },\label{waldtemprotES}
\end{eqnarray}
and hence the effective spatial dimensionality
$d_{\mbox{\tiny{eff}}}=D-2-\theta$. The mass and angular momentum of
the solution are given by
\begin{eqnarray}
M={\frac {\left( {\omega}^{2}+D-2-\theta\right)
\mbox{Vol}(\Sigma_{D-2}) }{2 \kappa\, \left( 1-{\omega}^{2} \right)
}}r_h^{D-1-\theta},\quad J= -{\frac {\left( D-1-\theta \right)
{\omega} \mbox{Vol}(\Sigma_{D-2}) }{2 \kappa\, \left( 1-{\omega}^{2}
\right)}}\,r_h^{D-1-\theta}.\label{momrotES}
\end{eqnarray}
On the other hand, the corresponding  soliton is described by the
following line element
\begin{eqnarray}
 d{s}^2 =\frac{1}{{r}^{\frac{2 \theta}{D-2}}}\left[-r^{2}\,{{d{t}}}^{2}+
 \frac{d{r}^{2}}{f({r})}+f({r})  \,d{{\varphi}^{2}}+
 r^2\,\sum_{i=1}^{D-3} d{{x}_{i}^{2}}\right],
 \label{metricdsoliton}
\end{eqnarray}
with the metric function and the scalar field given by
\begin{eqnarray*}\label{hvmrotatingESsoliton}
f({r})= r^2 \left\{1-\left[\frac{2 }{ (D-1-\theta)
\,{r}}\right]^{D-1-\theta}\right\},\quad \phi({r})=\sqrt {{\frac
{\theta\, \left( \theta-D+2 \right) }{\kappa(D-2)}}}\ln({r}).
\end{eqnarray*}
The mass of the soliton obtained through the quasilocal charge
expression (\ref{quasilocalcharge}) reads
\begin{eqnarray}
M_{\mbox{\tiny{sol}}}&=&-{\frac { \mbox{Vol}(\Sigma_{D-2})}{2
\kappa}}\,\left( {\frac {2}{ D-1-\theta}} \right)
^{D-1-\theta},\label{masssolES}
\end{eqnarray}
and it is straightforward to check that the Cardy-like formula
(\ref{CardyHvm}) with $d_{\mbox{\tiny{eff}}}=D-2-\theta$ fits
perfectly with the gravitational entropy (\ref{waldtemprotES}).

\subsection{Hyperscaling violation black hole with higher-order gravity theory}

As said before, the effective spatial dimensionality
$d_{\mbox{\tiny{eff}}}$ is not always equal to $D-2-\theta$ but may
have a different expression depending on the theory considered.
Nevertheless, in order to corroborate the Cardy-like formula with a
different value of the effective dimension, we opt for a pure
quadratic gravity theory defined by the action
\begin{eqnarray}
\frac{1}{2 \kappa}\,\int d^Dx\,\sqrt{-g}\left(\beta_1 R^2+\beta_2
R_{\mu\nu}R^{\mu\nu}\right),
\end{eqnarray}
with field equations given by
\begin{eqnarray}
\label{feqshvs} &&{\cal G}_{\mu\nu}:=\beta_2\square{R}_{\mu\nu}
+\frac12\left(4\beta_1+\beta_2\right)g_{\mu\nu}\square{R}
-\left(2\beta_1+\beta_2\right)\nabla_\mu\nabla_\nu{R}
+2\beta_2R_{\mu\alpha\nu\beta}R^{\alpha\beta}
+2\beta_1RR_{\mu\nu}\nonumber\\
&&-\frac12\left(\beta_1{R}^2+\beta_2{R}_{\alpha\beta}{R}^{\alpha\beta}
\right)g_{\mu\nu}=0.
\end{eqnarray}
After a straightforward computation, one can see that the field
equations admit the line element (\ref{rotmetricdhvm}) with
$\theta=D-1$ with the metric functions given by
\begin{eqnarray*}\label{hvmrotatingES}
N^2(r)&=& r^{2}
 {\left( 1-{\omega}^{2}
 \right)}\, F (r)
 \left( r^2-F(r) {\omega}^{2}\right) ^{-1}
, \quad R^2(r)=\frac{1}{1-{\omega}^{2}}\,\left(r^2- F
 \left( r \right) {\omega}^{2}\right)
,\\
N^{\varphi}(r)&=&\omega
\frac{\left(r^2-F(r)\right)}{\left(r^2-\omega^2 F(r)\right)},\quad
F(r)=r^2\Big(1-\left(\frac{r_h}{r}\right)^{\frac{2(D-1)}{D-2}}\Big),
\end{eqnarray*}
and, where the coupling constants are tied as
$$
\beta_1=-\frac{(D+2) \beta_2}{5D-2}.
$$
The entropy and temperature of the solution are given by
\begin{eqnarray}
\qquad {\cal S}=\frac{16\,\pi\,\mbox{Vol}(\Sigma_{D-2}) \left( D-1
 \right) ^{2}{\beta_2}}{ \left( 5\,D-2 \right)\left( D-2
 \right){\kappa} {\sqrt {1-{\omega}^{2}}}}\,r_h^{{\frac
{D}{D-2}}} ,\qquad T={\frac { \left( D-1 \right) \sqrt
{1-{\omega}^{2}}\,r_h}{ 2 \left( D-2 \right) \pi}} ,
\end{eqnarray}
which imply that the effective dimension is
$d_{\mbox{\tiny{eff}}}=D/(D-2)$. Without giving more details, we
just report the usual quantities of interest
\begin{eqnarray}
 M&=&\frac{4\,\left( D+{\omega}^{2} \left( D-2 \right)  \right) {\beta_2}
 \left( D-1 \right) ^{2}\mbox{Vol}(\Sigma_{D-2})}{
 \left( D-2 \right)^{2}{\kappa} \left( 5\,D-2 \right)
 \left( 1-{\omega}^{2} \right)}\,
r_h^{\frac {2(D-1)}{D-2}} ,\nonumber\\
J&=&-\frac{8\, \left( D-1 \right)
^{3}{\beta_2}\,\mbox{Vol}(\Sigma_{D-2})\omega}{ \left( D-2
\right)^{2}{\kappa}\left( 5\,D- 2 \right) \left( 1-{\omega}^{2}
\right)}\,r_h^{\frac
{2(D-1)}{D-2}},\nonumber\\
M_{\mbox{\tiny{sol}}}&=&-\frac{16 \mbox{Vol}(\Sigma_{D-2})\beta_2
}{(5\,D-2)\kappa}\,\left(\frac{D-2}{4}\right)^{\frac{D}{D-2}}\,
\left(\frac{4}{D-1}\right)^{\frac{2}{D-2}},
\end{eqnarray}
and again, we constat  the perfect matching between the
gravitational entropy and the Cardy-like formula (\ref{CardyHvm}).

\section{Testing the Cardy-like formula in the anisotropic case}
We now consider the anisotropic case which corresponds to a
dynamical exponent $z\not=1$ with our convention. In the static
case, the asymptotic metric of anisotropic (Lifshitz or hyperscaling
violating) black holes can be described by the following line
element
\begin{eqnarray}
ds^2=\frac{1}{r^{\frac{2\theta}{D-2}}}\Big[-r^{2z}dt^2+\frac{dr^2}{r^2}+r^2\sum_{i=1}^{D-2}dx_i^2\Big],
\end{eqnarray}
where now $z\not=1$ is responsible of the anisotropy between the
time and the space coordinates.

In spite of the fact that the Cardy-like formula (\ref{CardyHvm}) is
also appropriate with $z\not=1$, stationary anisotropic black hole
solutions are not known in the literature. Moreover, unlike the
isotropic case, the Lorentz boosts are not longer symmetries for
spacetimes with $z\not=1$, and hence the usual trick of performing a
Lorentz boost to the static solution may yield to a metric with a
rather obscure causal structure. These are the reasons for which we
will first concentrate on static  anisotropic black holes ($J=0$) in
order to test the consistency of the formulas
(\ref{CardyHvm}-\ref{CardyHvmelec}). Nevertheless, in the last
subsection, we will observe the effect on turning on the momentum of
a static Lifshitz black hole by the usual Lorentz transformation.
Making abstraction of the causal structure, we will compute the mass
and angular momentum of the resulting metric and see explicitly that
the Cardy-like formula with $J\not=0$ is still consistent with the
gravitational entropy.

\subsection{Lifshitz black holes with higher-order gravity theories\label{LifBH}}
We now deal with a gravity action in arbitrary dimension $D$ with
quadratic-curvature corrections given by
\begin{eqnarray}
S&=&\frac{1}{2\kappa}\int{d}^{D}x\sqrt{-g}
\left(R-2\Lambda+\beta_1{R}^2
+\beta_2{R}_{\alpha\beta}{R}^{\alpha\beta}
+\beta_3{R}_{\alpha\beta\mu\nu}{R}^{\alpha\beta\mu\nu} \right).
\label{eq:Squad}
\end{eqnarray}
The corresponding field equations read
\begin{eqnarray}
G_{\mu\nu}+\Lambda{g}_{\mu\nu}
+\left(\beta_2+4\beta_3\right)\square{R}_{\mu\nu}
+\frac12\left(4\beta_1+\beta_2\right)g_{\mu\nu}\square{R}
-\left(2\beta_1+\beta_2+2\beta_3\right)\nabla_\mu\nabla_\nu{R}
\nonumber\\ \nonumber\\
{}+2\beta_3R_{\mu\gamma\alpha\beta}R_{\nu}^{~\gamma\alpha\beta}
+2\left(\beta_2+2\beta_3\right)R_{\mu\alpha\nu\beta}R^{\alpha\beta}
{}-4\beta_3R_{\mu\alpha}R_{\nu}^{~\alpha}+2\beta_1RR_{\mu\nu}
\nonumber\\ \nonumber\\
{}-\frac12\left(\beta_1{R}^2+\beta_2{R}_{\alpha\beta}{R}^{\alpha\beta}
+\beta_3{R}_{\alpha\beta\gamma\delta}{R}^{\alpha\beta\gamma\delta}
\right)g_{\mu\nu}&=&0.\qquad \label{eq:squareGrav}
\end{eqnarray}
In Ref. \cite{AyonBeato:2010tm}, three families of Lifshitz black
hole solutions were found. In the present case, we are only
interested on the  family for which the dynamical exponent $z>
-(D-2)$ and described by the following line element\footnote{The two
remaining Lifshitz black hole solutions have a zero entropy.}
\begin{eqnarray}
ds^2=-r^{2z}\left[1-\left(\frac{r_h}{r}\right)^{\frac{z+D-2}{2}}\right]dt^2+
\frac{dr^2}{r^2\left[1-\left(\frac{r_h}{r}\right)^{\frac{z+D-2}{2}}\right]}+r^2\sum_{i=1}^{D-2}dx_i^2.
\end{eqnarray}
The coupling constants ensuring  the existence of this solution can
be found in \cite{AyonBeato:2010tm}. For this family of solution,
the entropy and temperature are
\begin{eqnarray}\label{entropytemlifshitz}
{\cal S}&=&-\frac{2\,\pi \, \mbox{Vol}(\Sigma_{D-2})}{\kappa
}\,{Q(z)r_{h} ^{D-2}},\label{waldentropybhso}
\\
T&=&\frac{(z+D-2) \left(r_{h}\right)^{z}}{8 \pi},
\end{eqnarray}
with
\begin{eqnarray}
Q(z)= \frac{\left(3\,{z}^{2}+(D-2)(D+2)\right)  \left( D-2+3\,z
\right) \left( D+2-3\,z \right)}{27z^4-4(27D-45)z^3
-(D-2)\big[2(5D-116)z^2+4(D^2-D+30)z+(D+2)(D-2)^2\big]}.\nonumber\\
\label{qz}
\end{eqnarray}
Since the field equations are of higher order, we find more
convenient to adopt the quasilocal formalism in order to compute the
mass. This will correspond to the charge $Q$ defined in
(\ref{quasilocalcharge}) with a Killing vector field
$\xi^t=\partial_t$. In the present case, the tensor  $P^{\alpha
\beta \gamma \delta}$ appearing in the charge formula
(\ref{quasilocalcharge}) is given by
\begin{eqnarray*}
P^{\alpha \beta \gamma \delta} &=&\frac{1}{4
\kappa}\,\Big(g^{\alpha\gamma}g^{\beta\delta}-g^{\alpha\delta}g^{\beta\gamma}\Big)
+\frac{\beta_{1}}{2\,\kappa}\, R \left(g^{\alpha \gamma } g^{\beta
\delta } -g^{\alpha \delta } g^{\beta \gamma
}\right)\nonumber\\
&+&\frac{\beta_{2}}{4\,\kappa}\, \left(g^{\beta \delta } R^{\alpha
\gamma }-g^{\beta \gamma } R^{\alpha \delta } -g^{\alpha \delta }
R^{\beta \gamma }+g^{\alpha \gamma } R^{\beta \delta
}\right)+\frac{\beta_3}{\kappa} R^{\alpha \beta \gamma \delta}.
\end{eqnarray*}
After a tedious but straightforward computation, one obtains the
expression of the mass
\begin{eqnarray}\label{masmomlifshitzsol1}
M=-\frac{(D-2)\,\mbox{Vol}(\Sigma_{D-2})}{4\kappa}{Q(z)\,r_{h}^{z+D-2}}.
\end{eqnarray}
As usual, the corresponding static soliton is
\begin{align}
d{s}^2 &= -{r}^{2} d {t}^2+ \frac{1}{{r}^2}\, \frac{d{r}^2}{f({r})}+
{r}^{2z}\,f({r})\,d
{\varphi}^{2}+{{r}^{2}}\sum_{i=1}^{D-3}\,d{x}_{i}^{2},\nonumber\\
f({r})&=1-\left[{\frac {4}{\left( z+D-2 \right)
}}\right]^{\frac{z+D-2}{2z}}\,\frac{1}{{r}^{\frac{z+D-2}{2}}},
\label{solitonlifshitz-cuad-corr}
\end{align}
and its mass is computed to be
\begin{eqnarray}
M_{\mbox{\tiny{sol}}}&=& {z\,\mbox{Vol}(\Sigma_{D-2})\,
\left[\frac{4}{\left(z+D-2\right)}\right]^{{\frac
{z+D-2}{z}}}}\,\frac{Q(z)}{4\kappa}.
\end{eqnarray}
It is interesting to note again that the expression of the entropy
(\ref{entropytemlifshitz}) coincides with the Cardy-like formula
with  $d_{\mbox{\tiny{eff}}}=D-2$, $J=0$ and for any value of the
dynamical exponent $z$.

\subsection{Charged anisotropic  black holes with two Abelian gauge fields}
In this subsection, we would like to check the charged version of
the Cardy-like formula (\ref{CardyHvmelec}) in the anisotropic and
static situation, $z\not=1$ and $J=0$. In order to achieve this
task, one considers the case of Einstein gravity with two abelian
fields $A_{(i)}$ and a dilaton $\phi$ with action
\begin{equation}
S=\frac{1}{2\kappa} \int d^Dx \sqrt{-g}\left( R - 2\Lambda -
\frac{1}{2}\partial_{\mu}\phi \partial^{\mu}\phi -
 \frac{1}{4} \sum_{i=1}^2 e^{\lambda_{i}\phi}\mathcal{F}_{(i)}^2 \right),
\end{equation}
with $\mathcal{F}_{(i)}^2 = F_{(i) \mu \nu} F_{(i)}^{\mu \nu}$ for
$i=1,2$. For the following ansatz
\begin{eqnarray}
ds^2 = -r^{2z}F(r) dt^2 + \frac{dr^2}{r^2 F(r)} + r^2 \sum_{i=1}^{D-2}dx_{i}^2, \\
A_{(i)\mu} dx^{\mu} = A_{(i)t} dt,\ \qquad \phi = \phi(r),\nonumber
\end{eqnarray}
a solution was found in \cite{Tarrio:2011de}
\begin{subequations}
\label{tarriosol}
\begin{eqnarray}
F(r) &=& 1 - m\left(\frac{r_h}{r}\right)^{z+D-2} + (m-1)\left(\frac{r_h}{r}\right)^{2(z+D-3)},\\
A_{(1)t} &=& \sqrt{\frac{2(z-1)}{z+D-2}}\mu^{-\frac{\lambda_1}{2}}(r^{z+D-2} - r_h^{z+D-2}), \\
A_{(2)t} &=& -\sqrt{\frac{2(m-1)(D-2)}{z+D-4}} \mu^{-\frac{\lambda_2}{2}}r_h^{z+D-3}(r^{-(z+D-4)} - r_h^{-(z+D-4)}), \\
e^{\phi} &=& \mu r^{\sqrt{2(D-2)(z-1)}}, \qquad \lambda_1 =
-\sqrt{\frac{2(D-2)}{z-1}},\qquad \lambda_2 =
\sqrt{\frac{2(z-1)}{D-2}},
\end{eqnarray}
\end{subequations}
where $m,\mu$ are integration constants, and $r_h$ stands for the
location of the horizon. Note that this presentation
(\ref{tarriosol}) is equivalent to the one considered in
\cite{Tarrio:2011de}, after some redefinitions of the constants. We
stick to (\ref{tarriosol}) for latter convenience. With our
notation, the Wald entropy and Hawking temperature read
\begin{eqnarray}
\label{tarrioent}{\cal S} &=& \frac{2\pi}{\kappa}r_h^{D-2} \mbox{Vol}(\Sigma_{D-2}), \\
T &=& \frac{\left[(z+D-4)(2-m)+2\right]}{4\pi}r_h^z,\,\nonumber
\end{eqnarray}
while the mass, electric potential and electric charge are
\begin{eqnarray*}
&&M =\frac{(D-2)m}{2\kappa}r_h^{z+D-2} \mbox{Vol}(\Sigma_{D-2}), \quad \Phi_e =\sqrt{\frac{2(D-2)(m-1)}{z+D-4}}\mu^{-\frac{\lambda_2}{2}} r_h,\\
\mathcal{Q}_e &=&
\frac{\sqrt{2(D-2)(m-1)(z+D-4)}\mu^{\frac{\lambda_2}{2}}}{2\kappa}
r_h^{z+D-3} \mbox{Vol}(\Sigma_{D-2}).
\end{eqnarray*}
It remains to derive the soliton counterpart from the uncharged
black hole solution which corresponds to the limit $m\to 1$. The
double Wick rotation takes the following form
\begin{eqnarray}
ds^2 &=& -r^2dt^2 + \frac{dr^2}{r^2 {f}(r)} + r^{2z}{f}(r)d\varphi^2 + r^2\sum_{i=1}^{D-3} dx_{i}^2,\\
\nonumber {f}(r) &=& 1 - \left(\frac{\tilde{r_h}}{r}\right)^{z+D-2},
\end{eqnarray}
where we have defined $$\tilde{r_h} =
\left(\frac{2}{z+D-2}\right)^{\frac{1}{z}}.$$ Using (\ref{eq:K}) ,
the variation of the Noether potential and the surface term read
\begin{eqnarray*}
\Delta K^{rt}= -\frac{\tilde{r_h}^{z+D-2}}{\kappa},\qquad \int_{0}^1
ds\ \Theta^{r} = -\frac{z-2}{2\kappa} \tilde{r_h}^{z+D-2},
\end{eqnarray*}
and then the mass of the soliton is
\begin{eqnarray}
M_{\mbox{\tiny{sol}}} =
-\frac{z\mbox{Vol}(\Sigma_{D-2})}{2\kappa}\left(\frac{2}{z+D-2}\right)^{\frac{z+D-2}{z}}.
\end{eqnarray}
It is now straightforward to check that the formula
(\ref{CardyHvmelec}) matches perfectly with the Wald entropy
(\ref{tarrioent}).

\subsection{Turning on the angular momentum}
We now turn on the angular momentum of the solution discussed in
Sec. \ref{LifBH} by operating a standard Lorentz transformation
$$
t\to \frac{1}{\sqrt{1-{\omega}^2}}(t+{\omega}\,\varphi),\qquad
\varphi\to \frac{1}{\sqrt{1-{\omega}^2}}(\varphi+{\omega} t).
$$
The resulting metric reads
\begin{eqnarray}\label{frotatinglifsol11}
 ds^2 &=&-N^2(r)
 dt^{2}+\frac{dr^{2}}{F(r)}+R^{2}(r)\,(d\varphi+N^{\varphi}(r)dt)^{2} +r^{2}
 \sum_{i=1}^{D-3} dx_{i}^{2},
\end{eqnarray}
 where
 \begin{eqnarray}\label{frotatinglifsol12}
N^2(r)&=& r ^{2(z+1)}H (r)
 \left( 1-{\omega}^{2} \right) \left[ r^{2}-
 r^{2\,z}H\left( r \right) {\omega}^{2}\right] ^{-1}
, \nonumber\\
R^2(r)&=&\frac{1}{(1-{\omega}^{2})}\,\left[{r}^{2}- r^{2\,z}H
 \left( r \right) {\omega}^{2}\right]
,\qquad F(r)=r^{2} H(r),\\
N^{\varphi}(r)&=&{\omega} \left[  {r}^{2}- {r}^{2\,z}\,H(r)
\right]\,\left[{r}^2-{r}^{2\,z}H
 \left( r \right) {\omega}^{2}
\right]^{-1},\qquad H(r)=
\left[1-\left(\frac{r_h}{r}\right)^{\frac{z+D-2}{2}}\right].\nonumber
\end{eqnarray}
As already mentioned, the resulting metric may suffer some pathology
essentially due to the fact that the combination
$r^2-r^{2z}H(r)\omega^2$ is not ensured to be positive for any value
of $r>0$ as it is the case in the isotropic situation $z=1$.
Nevertheless, making abstraction of this problem, one can still
compute the  entropy and temperature of the solution
\begin{eqnarray}\label{entropytemlifshitzrot}
{\cal S}&=&-\frac{2\,\pi \,
\mbox{Vol}(\Sigma_{D-2})}{\kappa\sqrt{1-\omega^2}}\,{Q(z)}\,
r_h^{D-2},\label{waldentropybhsol1}
\\
T&=&\frac{(z+D-2) \sqrt{1-\omega^{2}}r_h^{z}}{8 \pi},
\end{eqnarray}
where $Q(z)$ is defined in (\ref{qz}). One of the advantage of the
quasilocal formalism \cite{Kim:2013zha,Gim:2014nba} is precisely to
overcome the difficulty at infinity by introducing a one-parameter
and by integrating in the interior region and not at infinity. Since
the asymptotic form of the resulting metric
(\ref{frotatinglifsol11}-\ref{frotatinglifsol12}) is not clear, the
quasilocal formalism seems to be very-well appropriated to
circumvent this problem. In doing so, one can compute the mass and
the angular momentum
\begin{eqnarray}\label{masmomlifshitzrotsol1}
M&=&-\frac{(D-2+z\omega^{2})\,\mbox{Vol}(\Sigma_{D-2})}{4\kappa
(1-\omega^2)}
\,{Q(z)}\,r_h^{z+D-2}, \\
J&=&\frac{(z+D-2)\,\mbox{Vol}(\Sigma_{D-2}) \omega}{4\kappa
(1-\omega^2)}\,{Q(z)}\,r_h^{z+D-2}.
\end{eqnarray}

Finally, it is somehow appealing that the Cardy-like formula
(\ref{CardyHvm}) with the angular momentum turning on still
reproduces the correct value of the gravitational entropy.
\section{Summary and concluding remarks}

Here, we have considered rotating (an)isotropic black holes in
arbitrary dimension with a planar horizon which are obtained from
static configurations through a Lorentz transformation. The aim of
this paper is to show that the spinning black hole entropy can be
obtained from the microcanonical degeneracy of states according to a
Cardy-like formula making no reference to any central charge but
instead involving the mass of the ground state. The ground state is
in fact identified with a gravitational bulk soliton. Hence, one of
our working hypothesis in order to reproduce the semiclassical black
hole entropy is the existence of a soliton. From a technical
perspective, the soliton, in all the examples we have treated, is
obtained from the static black hole by a double analytic
continuation followed by a suitable rescaling that permits to absorb
the constant of integration. This procedure is quite similar to the
one that yields the AdS soliton \cite{Horowitz:1998ha}. In doing so,
the resulting solitonic solution turns to be smooth, regular and
devoid of any constant of integration fulfilling what the ground
state is expected to be. However, there exist black hole solutions
for which the double Wick rotation does not apply for different
reasons. For example, this can occur for  black holes for which the
topology of the event horizon presents an anisotropic scaling
symmetry. Such examples have been known much before the advent of
Lifshitz spacetimes \cite{Cadeau:2000tj}. In this reference, two
families of static black holes solutions of Einstein equations in
five dimensions with a negative cosmological constant were
constructed, and the horizon topologies of these solutions are
modeled by the Solv $3-$geometry and the Nil $3-$geometry. These
geometries are two of the eight geometries of the Thurston
classification. The Solv (resp. the Nil) solution is asymptotically
AdS (resp. Lifshitz with $z=3/2$) but both solution enjoys  an
anisotropy along one of the coordinates of the event horizon
responsible of the violation of the hyperscaling property. On one
hand, a simple calculation shows that the Cardy-Verlinde formula
\cite{Verlinde:2000wg} for the Solv solution does not yield the
correct temperature dependence. On the other hand, while the Solv
solution fits perfectly our assumptions, some complications have
emerged concerning the Nil solution, in particular to construct the
corresponding soliton. An interesting task will consist in
understanding what would be the soliton configuration for the Nil
solution or how to construct it (even numerically). One can go
further extending the analysis done in this paper to the many
examples of black holes with Thurston horizon topology.

Another aspect that may deserve some attention in the future has to
do with the Smarr formulas. These latter are relations expressing
the mass as a simple bilinear form involving the other conserved
charges and the thermodynamical quantities \cite{Smarr:1972kt}.
Smarr relations can also be viewed as the integral forms of the
first law of thermodynamics. For example, in all the cases studied
in this paper, the solutions satisfy a Smarr relation given
generically by
$$
M=\left(\frac{d_{\mbox{\tiny{eff}}}}{d_{\mbox{\tiny{eff}}}+z}\right)\,
T{\cal S}+\Omega J.
$$
In the case of asymptotically AdS black holes, extended versions of
the first law and of the Smarr formula have been obtained where the
cosmological constant is considered as a thermodynamic variable, see
e. g. \cite{Smarr}. In this perspective, the mass of the AdS black
hole may be understood as the enthalpy of spacetime while the
cosmological constant plays the role of a pressure term in the first
law. Recently, these ideas have been shown to hold also for Lifshitz
black holes \cite{Brenna:2015pqa}. Since Smarr and Cardy formulas
are intimately linked, it will be interesting to identify the
physical implications on the Cardy-like formulas of viewing the mass
as enthalpy.

Finally, we expect that the survey operating in this paper, apart
from confirming the validity of the Cardy-like formulas, will be of
relevance in order to clarify some issues concerning the field
theory side.

\begin{acknowledgments}
We thank Eloy Ay\'on-Beato for useful discussions. This work has
been partially supported by grant 1130423 from FONDECYT. MB is
supported by ``Plan de Mejoramiento Institucional'' UCM1310,
MINEDUC, Chile.
\end{acknowledgments}


\begin{thebibliography}{99}


\bibitem{Bekenstein:1973ur}
  J.~D.~Bekenstein,
  Phys.\ Rev.\ D {\bf 7}, 2333 (1973).
  doi:10.1103/PhysRevD.7.2333


\bibitem{Hawking:1974sw}
  S.~W.~Hawking,
  Commun.\ Math.\ Phys.\  {\bf 43}, 199 (1975)
  Erratum: [Commun.\ Math.\ Phys.\  {\bf 46}, 206 (1976)].
  doi:10.1007/BF02345020


\bibitem{Brown:1986nw}
  J.~D.~Brown and M.~Henneaux,
  Commun.\ Math.\ Phys.\  {\bf 104}, 207 (1986).
  doi:10.1007/BF01211590


\bibitem{Cardy:1986ie}
  J.~L.~Cardy,
  Nucl.\ Phys.\ B {\bf 270}, 186 (1986).
  doi:10.1016/0550-3213(86)90552-3


\bibitem{Banados:1992wn}
  M.~Banados, C.~Teitelboim and J.~Zanelli,
  Phys.\ Rev.\ Lett.\  {\bf 69}, 1849 (1992)
  doi:10.1103/PhysRevLett.69.1849
  [hep-th/9204099].



\bibitem{Strominger:1997eq}
  A.~Strominger,
  JHEP {\bf 9802}, 009 (1998)
  doi:10.1088/1126-6708/1998/02/009
  [hep-th/9712251].


\bibitem{Guica:2008mu}
  M.~Guica, T.~Hartman, W.~Song and A.~Strominger,
  Phys.\ Rev.\ D {\bf 80}, 124008 (2009)
  doi:10.1103/PhysRevD.80.124008
  [arXiv:0809.4266 [hep-th]].


 \bibitem{Azeyanagi:2009wf}
  T.~Azeyanagi, G.~Compere, N.~Ogawa, Y.~Tachikawa and S.~Terashima,
  Prog.\ Theor.\ Phys.\  {\bf 122}, 355 (2009)
  doi:10.1143/PTP.122.355
  [arXiv:0903.4176 [hep-th]].

  \bibitem{Verlinde:2000wg}
  E.~P.~Verlinde, {\it On the holographic principle in a radiation dominated universe},
  hep-th/0008140.


\bibitem{Detournay:2016gao}
  S.~Detournay, L.~A.~Douxchamps, G.~S.~Ng and C.~Zwikel,
  JHEP {\bf 1606}, 014 (2016)
  doi:10.1007/JHEP06(2016)014
  [arXiv:1602.09089 [hep-th]].

\bibitem{Gonzalez:2011nz}
  H.~A.~Gonzalez, D.~Tempo and R.~Troncoso,
  JHEP {\bf 1111}, 066 (2011)
  doi:10.1007/JHEP11(2011)066
  [arXiv:1107.3647 [hep-th]].


\bibitem{Shaghoulian:2015dwa}
  E.~Shaghoulian,
  JHEP {\bf 1511}, 081 (2015)
  doi:10.1007/JHEP11(2015)081
  [arXiv:1504.02094 [hep-th]].


\bibitem{Bravo-Gaete:2015wua}
  M.~Bravo-Gaete, S.~Gomez and M.~Hassaine,
  Phys.\ Rev.\ D {\bf 91}, no. 12, 124038 (2015)
  doi:10.1103/PhysRevD.91.124038
  [arXiv:1505.00702 [hep-th]] ; Phys.\ Rev.\ D {\bf 92}, no. 12, 124002 (2015)
  doi:10.1103/PhysRevD.92.124002
  [arXiv:1510.04084 [hep-th]].


\bibitem{Kachru:2008yh}
  S.~Kachru, X.~Liu and M.~Mulligan,
  Phys.\ Rev.\ D {\bf 78}, 106005 (2008)
  doi:10.1103/PhysRevD.78.106005
  [arXiv:0808.1725 [hep-th]].






  \bibitem{Correa:2010hf}
  F.~Correa, C.~Martinez and R.~Troncoso,
  JHEP {\bf 1101}, 034 (2011)
  doi:10.1007/JHEP01(2011)034
  [arXiv:1010.1259 [hep-th]].

 \bibitem{Correa:2011dt}
  F.~Correa, C.~Martinez and R.~Troncoso,
  JHEP {\bf 1202}, 136 (2012)
  doi:10.1007/JHEP02(2012)136
  [arXiv:1112.6198 [hep-th]].

\bibitem{Correa:2012rc}
  F.~Correa, A.~Faundez and C.~Martinez,
  Phys.\ Rev.\ D {\bf 87}, no. 2, 027502 (2013)
  doi:10.1103/PhysRevD.87.027502
  [arXiv:1211.4878 [hep-th]].





\bibitem{Horowitz:1998ha}
  G.~T.~Horowitz and R.~C.~Myers,
  Phys.\ Rev.\ D {\bf 59}, 026005 (1998)
  doi:10.1103/PhysRevD.59.026005
  [hep-th/9808079].



\bibitem{Ayon-Beato:2015jga}
  E.~Ayon-Beato, M.~Bravo-Gaete, F.~Correa, M.~Hassaine, M.~M.~Juarez-Aubry and J.~Oliva,
  Phys.\ Rev.\ D {\bf 91}, no. 6, 064006 (2015)
  doi:10.1103/PhysRevD.91.064006
  [arXiv:1501.01244 [gr-qc]].




\bibitem{Shaghoulian:2015lcn}
  E.~Shaghoulian,
  Phys.\ Rev.\ D {\bf 94}, no. 10, 104044 (2016)
  doi:10.1103/PhysRevD.94.104044
  [arXiv:1512.06855 [hep-th]].



\bibitem{Abbott:1981ff}
  L.~F.~Abbott and S.~Deser,
  Nucl.\ Phys.\ B {\bf 195}, 76 (1982);
  Phys.\ Lett.\ B {\bf 116}, 259 (1982);
  S.~Deser and B.~Tekin,
  Phys.\ Rev.\ Lett.\  {\bf 89}, 101101 (2002)
  [hep-th/0205318];
  Phys.\ Rev.\ D {\bf 67}, 084009 (2003)
  [hep-th/0212292];
  C.~Senturk, T.~C.~Sisman and B.~Tekin,
  Phys.\ Rev.\ D {\bf 86}, 124030 (2012)
  [arXiv:1209.2056 [hep-th]].



\bibitem{Kim:2013zha}
  W.~Kim, S.~Kulkarni and S.~H.~Yi,
  Phys.\ Rev.\ Lett.\  {\bf 111} (2013) no.8,  081101
   Erratum: [Phys.\ Rev.\ Lett.\  {\bf 112} (2014) no.7,  079902]
  doi:10.1103/PhysRevLett.112.079902, 10.1103/PhysRevLett.111.081101
  [arXiv:1306.2138 [hep-th]].


\bibitem{Gim:2014nba}
  Y.~Gim, W.~Kim and S.~H.~Yi,
  JHEP {\bf 1407} (2014) 002
  doi:10.1007/JHEP07(2014)002
  [arXiv:1403.4704 [hep-th]].

\bibitem{Lemos:1994xp}
  J.~P.~S.~Lemos,
  Phys.\ Lett.\ B {\bf 353}, 46 (1995)
  doi:10.1016/0370-2693(95)00533-Q
  [gr-qc/9404041].



\bibitem{Awad:2002cz}
  A.~M.~Awad,
  Class.\ Quant.\ Grav.\  {\bf 20}, 2827 (2003)
  doi:10.1088/0264-9381/20/13/327
  [hep-th/0209238].


\bibitem{Martinez:1999qi}
  C.~Martinez, C.~Teitelboim and J.~Zanelli,
  Phys.\ Rev.\ D {\bf 61}, 104013 (2000)
  doi:10.1103/PhysRevD.61.104013
  [hep-th/9912259].




\bibitem{Ayon-Beato:2015ada}
  E.~Ayon-Beato, M.~Hassaine and J.~A.~Mendez-Zavaleta,
  Phys.\ Rev.\ D {\bf 92}, no. 2, 024048 (2015)
  doi:10.1103/PhysRevD.92.024048
  [arXiv:1506.02277 [hep-th]].

\bibitem{Martinez:1996gn}
  C.~Martinez and J.~Zanelli,
  Phys.\ Rev.\ D {\bf 54}, 3830 (1996)
  doi:10.1103/PhysRevD.54.3830
  [gr-qc/9604021].


\bibitem{Henneaux:2002wm}
  M.~Henneaux, C.~Martinez, R.~Troncoso and J.~Zanelli,
  Phys.\ Rev.\ D {\bf 65}, 104007 (2002)
  doi:10.1103/PhysRevD.65.104007
  [hep-th/0201170].

\bibitem{LOV} D. Lovelock, ``The Einstein tensor and its generalizations,'' J. Math. Phys {\bf 12}, 498 (1971).



















\bibitem{CSsugra} M. Hassaine and J. Zanelli,
{\it Chern-Simons (super)gravity}, World Scientific.

\bibitem{Crisostomo:2000bb}
  J.~Crisostomo, R.~Troncoso and J.~Zanelli, Phys.\ Rev.\ D {\bf 62}, 084013 (2000)
  [hep-th/0003271].

 \bibitem{Boulware:1985wk}
  D.~G.~Boulware and S.~Deser,
  ``String Generated Gravity Models,''
  Phys.\ Rev.\ Lett.\  {\bf 55}, 2656 (1985).


\bibitem{Banados:1993ur}
  M.~Banados, C.~Teitelboim and J.~Zanelli,
  ``Dimensionally continued black holes,''
  Phys.\ Rev.\ D {\bf 49}, 975 (1994)
  [gr-qc/9307033].


\bibitem{Cai:1998vy}
  R.~-G.~Cai and K.~-S.~Soh,
  ``Topological black holes in the dimensionally continued gravity,''
  Phys.\ Rev.\ D {\bf 59}, 044013 (1999)
  [gr-qc/9808067].


\bibitem{Aros:2000ij}
  R.~Aros, R.~Troncoso and J.~Zanelli,
  ``Black holes with topologically nontrivial AdS asymptotics,''
  Phys.\ Rev.\ D {\bf 63}, 084015 (2001)
  [hep-th/0011097].



\bibitem{Cai:2001dz}
  R.~-G.~Cai,
"Gauss-Bonnet black holes in AdS spaces,''
  Phys.\ Rev.\ D {\bf 65}, 084014 (2002)
  [hep-th/0109133].


\bibitem{Cai:2001jc}
  R.~G.~Cai,
  Phys.\ Rev.\ D {\bf 63}, 124018 (2001)
  doi:10.1103/PhysRevD.63.124018
  [hep-th/0102113].

\bibitem{hvBH}
M.~Cadoni and M.~Serra,  JHEP {\bf 1211}, 136 (2012) ;
M.~Alishahiha, E.~O Colgain and H.~Yavartanoo,
  JHEP {\bf 1211}, 137 (2012); P.~Bueno, W.~Chemissany, P.~Meessen, T.~Ortin and C.~S.~Shahbazi,
  JHEP {\bf 1301}, 189 (2013)
  [arXiv:1209.4047 [hep-th]]; M.~Hassaine,
  Phys.\ Rev.\ D {\bf 91}, no. 8, 084054 (2015)






\bibitem{Dong:2012se}
  X.~Dong, S.~Harrison, S.~Kachru, G.~Torroba and H.~Wang, JHEP {\bf 1206}, 041 (2012).



\bibitem{Alishahiha:2012cm}
  M.~Alishahiha and H.~Yavartanoo, JHEP {\bf 1211}, 034 (2012).








\bibitem{Perlmutter:2012he}
  E.~Perlmutter,
  JHEP {\bf 1206}, 165 (2012).



\bibitem{AyonBeato:2010tm}
  E.~Ayon-Beato, A.~Garbarz, G.~Giribet and M.~Hassaine, Phys.\ Rev.\ D {\bf 80}, 104029 (2009)
  doi:10.1103/PhysRevD.80.104029
  [arXiv:0909.1347 [hep-th]] ; JHEP {\bf 1004}, 030 (2010)
  [arXiv:1001.2361 [hep-th]].

 \bibitem{Tarrio:2011de}
  J.~Tarrio and S.~Vandoren,
  JHEP {\bf 1109}, 017 (2011)
  doi:10.1007/JHEP09(2011)017
  [arXiv:1105.6335 [hep-th]].















\bibitem{Cadeau:2000tj}
  C.~Cadeau and E.~Woolgar,
  Class.\ Quant.\ Grav.\  {\bf 18}, 527 (2001)
  doi:10.1088/0264-9381/18/3/312
  [gr-qc/0011029].



\bibitem{Smarr:1972kt}
  L.~Smarr,
  Phys.\ Rev.\ Lett.\  {\bf 30}, 71 (1973)
  Erratum: [Phys.\ Rev.\ Lett.\  {\bf 30}, 521 (1973)].
  doi:10.1103/PhysRevLett.30.71



\bibitem{Smarr}
 M.~M.~Caldarelli, G.~Cognola and D.~Klemm,
  Class.\ Quant.\ Grav.\  {\bf 17}, 399 (2000)
  doi:10.1088/0264-9381/17/2/310;
  D.~Kastor, S.~Ray and J.~Traschen,
  Class.\ Quant.\ Grav.\  {\bf 26}, 195011 (2009)
  doi:10.1088/0264-9381/26/19/195011
  [arXiv:0904.2765 [hep-th]].


\bibitem{Brenna:2015pqa}
  W.~G.~Brenna, R.~B.~Mann and M.~Park,
  Phys.\ Rev.\ D {\bf 92}, no. 4, 044015 (2015)
  doi:10.1103/PhysRevD.92.044015
  [arXiv:1505.06331 [hep-th]].
\end{thebibliography}
\end{document}